\documentclass[journal]{IEEEtran}

\ifCLASSINFOpdf
\else
\fi

\usepackage{todonotes}
\usepackage{amsmath,amsfonts,epsfig,graphicx}
\usepackage{subfig}
\usepackage{color,soul}
\usepackage{epstopdf}
\usepackage{tikz}
\usepackage{balance}
\usepackage{url}
\usepackage{multirow,hhline}
\usepackage{placeins}
\usepackage{wasysym}
\usepackage{amssymb}
\usepackage{pifont}
\usepackage[nomessages]{fp}
\usepackage{pgfplots}
\pgfplotsset{compat=newest}
\usetikzlibrary{plotmarks}
\usetikzlibrary{matrix}

\newcommand{\ru} {\rule{0mm}{3mm}}

\newcommand{\hM} {\widehat{M}}
\newcommand{\hMd}{\widehat{M}_\downarrow}
\newcommand{\hMa}{\widehat{M}_{{\downarrow}{a}}}
\newcommand{\tM} {\widetilde{M}}
\newcommand{\Plp}{P^{\rm lp}}
\newcommand{\hMlp} {\hM^{\rm lp}}

\newcommand{\LL}{{\cal L}}

\newcommand{\DL} {D^{_{\rm (K)}}_\lambda}
\newcommand{\DLa}{D^{_{\rm (K)}}_{\lambda, \rm align}}
\newcommand{\DR} {D_{\rho}}
\newcommand{\DS} {D_{S}}
\newcommand{\DSR}{D^{_{\rm (R)}}_S}
\newcommand{\RERGAS}{{\scriptsize R-ERGAS}}
\newcommand{\LPNN}  {$\lambda$-PNN}

\begin{document}

\title{Unsupervised Deep Learning-based Pansharpening with Jointly-Enhanced Spectral and Spatial Fidelity}

\author{
Matteo~Ciotola, Giovanni~Poggi, Giuseppe~Scarpa
\thanks{
M.Ciotola and G.Poggi are with the Department of Electrical Engineering and Information Technology, University Federico II, Naples, Italy, e-mail: \{firstname.lastname\}@unina.it.
G.Scarpa is with the Engineering Department of the University Parthenope, Naples, Italy, e-mail: giuseppe.scarpa@uniparthenope.it.}
}
\maketitle

\begin{abstract}
In latest years, deep learning has gained a leading role in the pansharpening of multiresolution images.
Given the lack of ground truth data, most deep learning-based methods carry out supervised training in a reduced-resolution domain.
However, models trained on downsized images tend to perform poorly on high-resolution target images.
For this reason, several research groups are now turning to unsupervised training in the full-resolution domain,
through the definition of appropriate loss functions and training paradigms.
In this context, we have recently proposed a full-resolution training framework which can be applied to many existing architectures.

Here, we propose a new deep learning-based pansharpening model that fully exploits the potential of this approach and provides cutting-edge performance.
Besides architectural improvements with respect to previous work, such as the use of residual attention modules,
the proposed model features a novel loss function that jointly promotes the spectral and spatial quality of the pansharpened data.
In addition, thanks to a new fine-tuning strategy, it improves inference-time adaptation to target images.
Experiments on a large variety of test images, performed in challenging scenarios,
demonstrate that the proposed method compares favorably with the state of the art both in terms of numerical results and visual output.
Code is available online at https://github.com/matciotola/Lambda-PNN.
\end{abstract}

\IEEEpeerreviewmaketitle

\section{Introduction}

Due to physical and technological constraints,
passive sensors are subject to a trade-off between spectral and spatial resolution: the more spectral bands, the lower their spatial resolution.
To overcome this limitation,
multi-resolution observation systems mount two sensors on the same flying platform,
a multispectral (MS) sensor with high spectral resolution and a single-band panchromatic (PAN) sensor with high spatial resolution.
The two pieces of information are then fused off-line through the so-called pansharpening process \cite{Vivone2015}
to recover the ideal full-resolution image.

Pansharpening has been a very active research field for more than twenty years.
With the advances in technology and the increasing diffusion and use of remote sensing multi-resolution data,
a growing number of researchers have confronted the problem of pansharpening, proposing several ingenious approaches.
According to the taxonomy proposed in \cite{Vivone2020}, these can be roughly cast into four main categories:
component substitution (CS) \cite{Shettigara1992, Laben2000, Aiazzi2007, Garzelli2008, Choi2011, Garzelli2015, Vivone2019},
multiresolution analysis (MRA) \cite{Ranchin2000, Aiazzi2002, Otazu2005, Alparone2017, Lolli2017, Vivone2018, Vivone2018a},
variational optimization (VO) \cite{Vivone2015a, Vicinanza2015, Palsson2014}, and
machine/deep learning (ML/DL) \cite{Masi2016, Yang2017, Wei2017L, Masi2017, Yuan2018, Zhan2019, He2019, Vitale2018, Deng2020}.
Both CS and MRA methods inject the PAN component into the resized (by interpolation) MS image, using however different schemes.
The former perform the injection in a transform domain,
obtained for example through principal component analysis (PCA) \cite{Chavez1989}, Gram-Schmidt (GS) projection \cite{Laben2000}, generalized Intensity-Hue-Saturation (G-IHS) transform \cite{Tu2004},
where the ``strongest'' component is replaced by a suitably equalized version of the PAN image.
The latter operate on ``detail'' (high-frequency) components,
hence requiring multi-resolution decomposition transforms such as Wavelets \cite{Nunez1999, Ranchin2000, Otazu2005, Khan2008} or Laplacian pyramids \cite{Aiazzi2002, Aiazzi2003, Aiazzi2006, Lee2010, Restaino2017}.
VO methods, instead, rely on the optimization of suitable acquisition or representation models,
such as resolution downgrading models \cite{Vivone2015a}, sparse representations \cite{Vicinanza2015}, total variation \cite{Palsson2014}, and low-rank matrix representations \cite{Palsson2020}.

In recent years, deep learning-based methods have taken center stage.
The impressive performance observed in a number of computer vision and image processing problems,
from denoising to super-resolution, to detection and classification,
raised high expectations for similar improvements in remote sensing applications.
However, for pansharpening, this quantum leap is yet to come,
even though the first deep learning-based method, PNN \cite{Masi2016}, dates back to 2016,
and a plethora of other solutions have been proposed since then
\cite{Yang2017, Wei2017a, Wei2017L, Rao2017, Masi2017, Azarang2017, Yuan2018, Liu2018, Shao2018, Vitale2018, Zhan2019,Dong2021A,Dong2021B,Gong2022,Gong2023}.
Two peculiar issues of pansharpening have long prevented reaping the full potential of deep learning:
poor generalization ability and lack of ground truth data.
Let us briefly analyze these problems.

In remote sensing,
the wild variability of observed images, due to diversity of sensors, scenes, and operating conditions,
make it difficult to generalize to data not seen in training.
In computer vision, this problem is usually solved by increasing the training set and by using suitable forms of augmentation.
Such solutions are hardly viable in remote sensing,
due to the scarcity of high-quality training data (often proprietary) and the peculiarities of multiresolution imaging.
Only recently, some field-specific forms of augmentation have been proposed \cite{Chen2023}
which appear to provide appreciable improvements.
A more radical solution was proposed in \cite{Scarpa2018a}.
The available pre-trained model is adapted to the statistics of the target image by a few cycles of fine tuning,
trading off some processing time for a largely improved quality.
This approach, called target-adaptive operating modality, solves the generalization issue at its root
and is rapidly gaining popularity.

Turning to the second limiting issue,
the lack of ground truth data prevents the use of straightforward and effective supervised training.
To cope with this problem,
early proposals \cite{Masi2016, Yang2017} resorted to supervised training in a reduced resolution domain.
The network was trained on decimated versions of PAN and MS components, with the original MS acting as ground truth.
The underlying assumption was that networks trained at low resolution
would work equally well in the full-resolution domain.
Experimental evidence, however, has clearly shown this scale-invariance hypothesis not to hold.
Good performance at low resolution hardly scales to full resolution, and low-quality images are often obtained.
Therefore, the idea of training pansharpening models on the original high-resolution data has lately gained momentum \cite{Luo2020, Ciotola2022}.
With this approach, there is no scale mismatch problem.
On the downside, lacking a ground truth, one must forsake supervised training and turn to alternative unsupervised solutions
\cite{Luo2020, Ciotola2022, Uezato2020, Ma2020, Seo2020}.
In particular, in \cite{Ciotola2022} we developed a full-resolution training framework based on {\it ad hoc} losses,
such that the quality of the fused image is assessed indirectly,
by measuring how closely it matches the PAN in the spatial domain and the MS in the spectral domain.
Early experiments show clear quality improvements for the resulting pansharpened images, fully supporting the validity of this choice.

This short analysis highlights recent important theoretical and practical advances in deep learning-based pansharpening.
Building upon these ideas and tools,
we propose here a new deep learning-based pansharpening method which further improves on the current state of the art.
First of all, we operate at full resolution, avoiding any processing of the original data that could impair image quality.
We start from the full-resolution training framework developed in \cite{Ciotola2022} and improve it in two ways:
{\it  i)} by adopting a new spectral loss, with perceptually motivated reprojection metrics in place of Euclidean norms; and
{\it ii)} by improving the accuracy of such metrics through loss-time re-alignment of spectral bands.
Then, in the context of this improved training framework,
we propose a new architecture which takes advantage of several promising and established tools,
such as residual learning, and spatial and channel attention modules.
To ensure good generalization to new data, we adopt the target-adaptive operating modality.
However, we develop a new data pre-processing scheme which removes its computational bottlenecks,
thus enabling its use in all real-world applications with off-the-shelf hardware.

We carried out a large number of experiments to validate our design choices
and to assess the performance of the proposed method in comparison with state-of-the-art competitors, both model-based and data-driven.
To this end, we considered a wide variety of datasets and the most challenging cross-dataset operating conditions.
Numerical results show the proposed method to be always among the best performers, and often the best.
Output images are characterized by high spatial and spectral fidelity,
displaying a fine PAN-MS co-registration, obtained with no intervention on the part of the user.
Our software tool is available online (upon publication) to enable in-depth analysis and further developments.

In summary, the main contributions of this work are
\begin{itemize}
\item   novel attention-based residual learning architecture;
\item   full-resolution training with perceptually-motivated spectral and spatial losses;
\item   automatic PAN-MS co-registration at sub-pixel precision;
\item   self-contained software module available online characterized by fast (almost size invariant) processing;
\item   thorough experimental validation on a large number of datasets and in challenging conditions.
\end{itemize}

In the rest of the manuscript we
account for related work (Section II),
describe in detail the proposed method (Section III),
carry out ablation studies
and
report on comparative experimental results (Section IV),
and, finally, draw conclusions (Section V).

\section{Related work}

In this Section, we review the state of the art on deep learning-based pansharpening.
However, we neglect methods trained at reduced resolution, referring the reader to a recent review \cite{Deng2022},
and focus on those performing unsupervised training in the full resolution domain, more strictly related to our proposal.
In addition, we describe the correlation-based spatial loss proposed in \cite{Ciotola2022} and adopted here.

\subsection{Unsupervised methods in the full resolution domain.}
Methods that work at full resolution overcome the scale mismatch problem, opening the way to superior performance.
Lacking a ground truth,
the central problem becomes the definition of a suitable unsupervised loss function
which provide meaningful indications for network optimization.
To the best of our knowledge,
only four very recent papers \cite{Luo2020, Uezato2020, Ma2020, Ciotola2022} have been proposed to date on this topic.
All of them define the loss function as a weighted sum of spectral and spatial (sometimes called structural) terms, $\LL_\lambda$, and $\LL_S$,
plus some possible additional terms, accounting for hybrid features, adversarial games, or simple weight regularization.
Tab.\ref{tab:unsup-losses} provides a synoptic view of such unsupervised losses,
while Tab.\ref{tab:Symbols} lists the domain-specific symbols most frequently used in the following of the paper.

\newcommand{\myrule}[1]{{ $\rule{0mm}{#1}_{\rule{0mm}{#1}}$ }}
\newcommand{\myname}[1]{{ \begin{minipage}{18mm} \center{#1} \end{minipage} }}
\newcommand{\myspec}[1]{{ \begin{minipage}{45mm} \center{#1} \end{minipage} }}
\newcommand{\myspat}[1]{{ \begin{minipage}{36mm} \center{#1} \end{minipage} }}
\newcommand{\mynote}[1]{{ \begin{minipage}{54mm} {#1} \end{minipage} }}
\newcommand{\norm}[1]{{\left\| #1 \right\|}}

\begin{table*}
\centering
\begin{tabular}{cccc} \hline
\myrule{4mm}
\myname{Loss name}     &
\myspec{$\LL_\lambda$} &
\myspat{$\LL_S$}       &
\mynote{more details}  \\ \hline

\myrule{6mm}
\myname{$\LL_{\rm SSQ}$ \cite{Luo2020}}                         &
\myspec{$\norm{\hMlp - \tM}_2 + [1-\text{SSIM}(\hMlp,\tM)]$}    &
\myspat{$\left\|P-I\right\|_2 + [1-\text{SSIM}(P, I)]$}         &
\mynote{$\{\alpha_b\}$ estimated at reduced resolution;         \\
additional regularization term $\LL_{\rm reg}$}             \ru \\ \hline

\myrule{6mm}
\myname{$\LL_{\rm GDD}$ \cite{Uezato2020}}                      &
\myspec{$\norm{\hMd - M}_2$}                                    &
\myspat{$\norm{\nabla P - \nabla I}_1$}                         &
\mynote{$\{\alpha_b\}$ learned in the training phase;           \\
$\hMd$: bicubic downscaling of $\hM$.}                      \ru \\ \hline

\myrule{8mm}
\myname{$\LL_{\rm PG}$ \cite{Ma2020}}                           &
\myspec{$\norm{\hMd - M }_2$}                                   &
\myspat{$\norm{\nabla P - \nabla I }_2$}                        &
\mynote{$\alpha_b=1/B \; \forall b \to I=\langle\hM\rangle$;    \\
$\hMd$: bilinear downscaling of $\hM$;                      \ru \\
additional adversarial term $\LL_{\rm adv}$.}               \ru \\ \hline

\myrule{8mm}
\myname{$\LL_{\rm Z-PNN}$ \cite{Ciotola2022}}                                               &
\myspec{$\norm{\hMd - M }_1$}                                                               &
\myspat{$\left\langle (1-\rho^{\sigma}) {\rm u}(\rho^{\max}-\rho^{\sigma}) \right\rangle$ } &
\mynote{$\rho^\sigma = {\rm corr}(P,\hM), \; \rho^{\max} = {\rm corr}(P^{\rm lp},\tM)$;     \\
$\hMd$: low-pass filtering and decimation of $\hM$;                                         \\
${\rm u}(\cdot)$: unit-step function.}                                                  \ru \\ \hline
\end{tabular}
\caption{Unsupervised losses proposed in the literature for high-resolution pansharpening. $I = \sum_b \alpha_b \hM_b \simeq P$.}
\label{tab:unsup-losses}
\end{table*}

\begin{table}[t]
\centering
\begin{tabular}{cl}
\hline
\textbf{Symbol}            & \textbf{Description}\\\hline
$R$                        & resolution ratio \\
$B$                        & number of multispectral bands \\
$M$, $P$                   & original multispectral and panchromatic components \\
$\hM$                      & pansharpened image \\
$\hM_\downarrow,\hMa$      & ($R{\times}$) downscaled version of $\hM$ wo/with alignment \\
$\tM$                      & ($R{\times}$) upscaled version of $M$ \\
$\Plp, \hMlp$              & low-pass filtered versions of $P$ and $\hM$ \\
$\LL_\lambda, \LL_S, \LL$  & spectral, spatial and total loss \\
$\langle \cdot \rangle$    & spatial and/or spectral average \\
\hline
\end{tabular}
\caption{Main symbols and operators used in the paper.}
\label{tab:Symbols}
\end{table}

Let us first consider the spectral loss term, $\LL_\lambda$,
whose goal is to quantify the consistency of the fused output image, $\hM$,
with the $R{\times}R$ times smaller input MS, $M$.
To compare them, these two images must be adjusted to the same size,
which can be done by upscaling $M$ \cite{Luo2020}, or downscaling $\hM$ \cite{Uezato2020, Ma2020, Ciotola2022}.
In both cases, there are implementation choices that impact heavily on the final result.
Only Luo {\em et al.} \cite{Luo2020} follow the first path.
The MS is expanded, $M \to \widetilde{M}$, through interpolation (not specified in the paper).
Since this latter image lacks high-frequency details, also the model output is smoothed $\hM \to \hM^{\rm lp}$,
with a low-pass Gaussian-shaped filter. 
Finally, a combination of Euclidean distance and structural similarity index (SSIM) is used to compare them.
The other path requires downgrading the model output, $\hM \to \hMd$, so as to compare it with the original MS.
A clear advantage of this second solution is that the spectral reference $M$ is left unaltered.
On the other hand, the procedure used to downgrade $\hM$ also affects the accuracy of spectral consistency measurement.
In \cite{Uezato2020} and \cite{Ma2020}, standard interpolators are used before decimation, bicubic and bilinear, respectively.
In our previous work \cite{Ciotola2022}, instead,
a physically explainable process is implemented, with a Gaussian-shaped low-pass filter matching the modulation transfer function (MTF).
In all cases, $L_p$ norms are used to compute the loss,
Euclidean in \cite{Uezato2020} and \cite{Ma2020},
$L_1$ in \cite{Ciotola2022} to favour sharper edges and ensure faster convergence.

The goal of the spatial loss, $\LL_S$, instead,
is to measure how accurately the spatial structures of the high-resolution PAN are reproduced in the pansharpened image.
\cite{Luo2020, Uezato2020} and \cite{Ma2020} all make the fundamental assumption
that the PAN can be accurately estimated through a linear combination of the high-resolution spectral bands
\begin{equation}
    I = \sum_{b=1}^B \alpha_b \hM_b \simeq P
    \label{eq:ms_to_pan}
\end{equation}
In other words, a space-invariant linear relationship is assumed to exist between the MS and PAN domains.
Given this assumption,
the problem reduces to computing the regression coefficient $\{\alpha_b\}$.
They are estimated on a reduced resolution dataset in \cite{Luo2020},
learned during the training process in \cite{Uezato2020}, and simply assumed to be all equal in \cite{Ma2020}.
The spatial loss is then computed by measuring the dissimilarity between $I$ and $P$.
In \cite{Luo2020} a combination of Euclidean distance and SSIM is again proposed,
while both \cite{Uezato2020} and \cite{Ma2020}, to better preserve high-frequency details,
work on the gradients of $I$ and $P$, using $L_1$ and $L_2$ norm, respectively.

Considering the limits and risks of a linear space-invariant model, already pointed out in \cite{Thomas2008},
in \cite{Ciotola2022} we propose a radically different correlation-based spatial loss.
Given its importance for this work, we describe it in detail below.

\subsection{Correlation-based spatial loss}

\newcommand{\wM}{{\widehat{M}}}
Let $X$ and $Y$ be two image patches, and let $\sigma^2_X$, $\sigma^2_Y$ and $\sigma_{XY}$ indicate their sample variances and covariance.
Then, the correlation coefficient between $X$ and $Y$ is defined as
\begin{equation}
    \rho_{XY} = {\rm Corr(X,Y)} = \frac{\sigma_{XY}}{\sigma_X\sigma_Y}, \hspace{6mm} -1 \leq \rho_{XY} \leq 1
    \label{eq:correlation}
\end{equation}
The correlation coefficient indicates to what extent one image can be linearly predicted from the other,
with $|\rho|=1$ implying perfect predictability and $\rho=0$ total incorrelation.

Now, the role of the spatial loss is to inject the detailed spatial structures of the PAN into the output image.
Accordingly, one may be tempted to define a loss that forces the correlation between the PAN and each pansharpened band to be unitary.
However, this would produce exact replicas of the PAN in each band, which is not our goal.
We want to preserve the peculiar dynamics of each band, tolerating also the presence of local areas with low correlation.
Therefore, we apply the basic idea with two corrections:
{\it  i)} only {\em local} correlations are considered;
{\it ii)} they are not forced to be unitary but only to reach a suitable reference level.

More formally, let
$\hM^\sigma(i,j,b)$ be a $\sigma\times\sigma$ patch drawn from band $b$ of the pansharpened image at spatial location $(i,j)$, and
$P^\sigma(i,j)$ the corresponding patch drawn from the PAN.
We compute the correlation field
\begin{equation}
    \rho^\sigma(i,j,b) = {\rm Corr}(P^\sigma(i,j),\hM^\sigma(i,j,b))
\end{equation}
At the same time, from low-pass versions of the same quantities, $P^{\rm lp}$ and $\widetilde{M}$,
we compute a reference correlation field, $\rho^{\sigma,\max}(i,j,b)$, which represents the local target for the correlation (the reader is referred to \cite{Ciotola2022} for more details).
Eventually, we define the local loss as
\begin{equation}
    \ell^\sigma(i,j,b) = \left\{ \begin{array}{ll}
            1-\rho^\sigma(i,j,b) & \rho^\sigma < \rho^{\sigma,\max} \\
            0                    & \mbox{otherwise}
            \end{array} \right.
\end{equation}
and the overall spatial loss as its average on space and bands
\begin{equation}
    \LL_S = \langle \ell^\sigma(i,j,b) \rangle
    \label{eq:ls-mean}
\end{equation}

In practice, this loss pushes the local correlation to grow, but only until the conservative target value defined by the reference field is reached.
Clearly, the size of the correlation window, $\sigma$, plays a critical role in performance.
It should be large enough to allow reliable estimates over the local window, but not so large to lose resolution.
Experiments in \cite{Ciotola2022} show $\sigma=R$, the PAN-MS resolution ratio, to be a good compromise value.
The reference field, instead, uses a window of size $R^2$, since it is computed on smoother images.

\section{Proposed Method}
\label{sec:proposed}

Our proposal builds upon the full resolution training framework proposed in \cite{Ciotola2022}.
Here, we introduce a new spectral loss which, thanks to accurate co-registration,
collaborates with the correlation-based spatial loss to provide enhanced spectral and spatial fidelity.
In addition, we propose a fast training adaptation procedure and a deeper architecture including residual and attention modules.

\subsection{Improved spectral loss}

The $L_1$ and $L_2$ norms often used in the spectral losses are simple and popular measures of distortion
but only loose proxies of the image quality as perceived by the end user.
Several perceptual metrics have been proposed over the years that better fit the human visual system,
such as the structural similarity (SSIM) \cite{Wang2004} or the universal image quality index (UIQI) \cite{Wang2002}.
In pansharpening, the use of $L_p$ norms as quality indicators has been questioned for a long time
and several alternative metrics have been proposed and are by now widely accepted.

\newcommand{\x}{{\bf x}}
\newcommand{\y}{{\bf y}}
A first popular measure of spectral quality is the {\em Erreur Relative Globale Adimensionnelle de Synth{\'e}se} (ERGAS) \cite{Wald2002}.
Given a $B$-band image, $\x$, and its approximation, $\y$, it is defined as
\begin{equation}
    {\rm ERGAS}(\y,\x) = \frac{100}{R} \sqrt{\frac{1}{B}\sum_{b=1}^B \frac{\|\y_b-\x_b\|^2}{\langle \x_b \rangle^2}}
\end{equation}
Also widespread is the $Q2^n$ metric \cite{Garzelli2009}.
It generalizes to $b$-bands images the UIQI metric defined for single-band images $\y$ and $\x$, as
\begin{equation}
    {\rm UIQI}(\y,\x) = \left\langle
                        \frac{\sigma_{xy}}{\sigma_x \sigma_y} \cdot
                        \frac{2\sigma_x\sigma_y}{\sigma_x^2 + \sigma_y^2} \cdot
                        \frac{2\mu_x \mu_y}{\mu_x^2 + \mu_y^2}
                        \right\rangle ,
\end{equation}
where $\mu_x$, $\mu_y$, $\sigma_x$, $\sigma_y$, $\sigma_{xy}$ are sample means, variances and covariance
computed over a $W\times W$ window centered on corresponding pixels $x$ and $y$ of $\x$ and $\y$, respectively.
To extend UIQI to the $B$-band images, all statistics are re-defined in vector/matrix form and properly combined \cite{Garzelli2009},
but the overall behavior of the metric remains the same.
In the remote sensing field, there is wide agreement on the meaningfulness of both these metrics.
Interestingly, they complement each other.
In fact, while ERGAS is based on pixel-wise measurements, $Q2^n$ relies on local statistics computed on relatively large windows.

Note that if only a reduced resolution target image, $\x'$, is available,
these metrics can still be used by properly downgrading the estimate $\y$ as well.
For example, the spectral distortion index $D^{\rm (K)}_\lambda$ proposed by Khan {\em et al.} \cite{Khan2009}
can be seen as an extension of $Q2^n$
\begin{equation}
    D^{\rm (K)}_\lambda (\y_\downarrow,\x') = 1 - Q2^n(\y_\downarrow, \x')
\end{equation}
Similar extensions have been also recently proposed \cite{Scarpa2022a} for ERGAS and other indexes.

Given the above considerations, we chose to define a new spectral loss based on these perceptually relevant metrics.
In particular, to exploit the complementary features exhibited by ERGAS and $D_\lambda^{(K)}$, we consider
their linear combination, with a weighting parameter, $\gamma$, to be set on the basis of experiments.
In more detail, we use these metrics to compare the original MS component with the pansharpened image reprojected in the MS domain.
This latter process consists in applying an MTF-matched Gaussian filter to all bands of $\hM$ and decimating them.
Eventually, the proposed spectral loss reads as
\begin{eqnarray}
    \LL_\lambda = {\DL}(\hM_\downarrow, M) + \gamma {\rm ERGAS}(\hM_\downarrow, M)
    \label{eq:ll}
\end{eqnarray}
Note that all terms are computed with respect to the original MS component, $M$,
which is neither co-registered nor expanded or resampled, all operations that could degrade the quality of the reference.

\subsection{Co-registration at loss}

\newcommand{\fns}{\footnotesize}
\newcommand{\scr}{\scriptsize}
\newcommand{\pathCoreg}{./png_coreg/}
\newcommand{\imCoreg}[1]{\includegraphics[width=3.6cm]{\pathCoreg #1.jpg}}
\begin{figure}
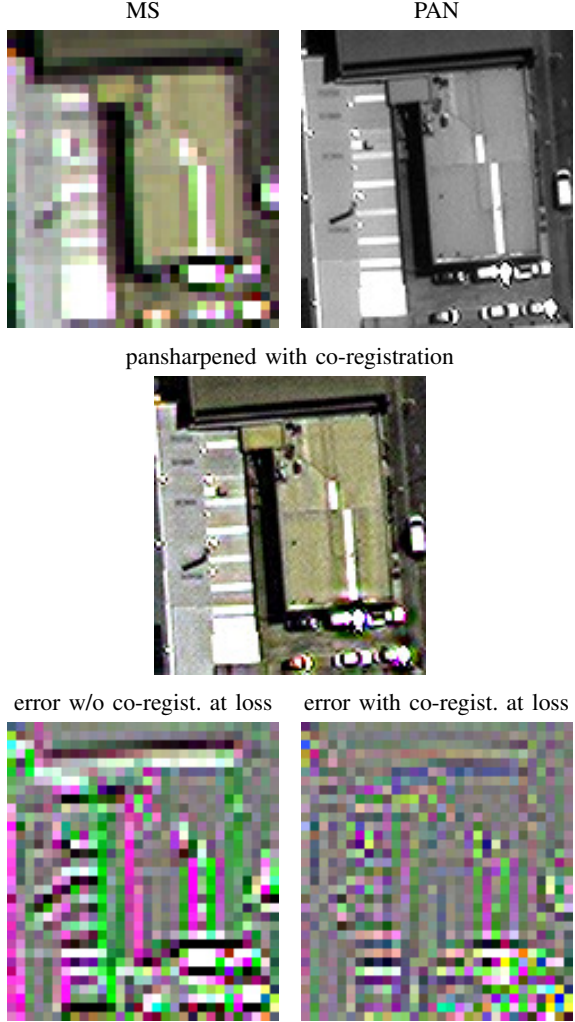

\centering
\tiny
\setlength{\tabcolsep}{1pt}
\begin{tabular}{c@{\hspace{3mm}}c}
\vspace{1mm} \small MS                            & \small PAN                            \\
\vspace{2mm} \imCoreg{Q_MS_1}                     & \imCoreg{Q_PAN_1}                     \\
\multicolumn{2}{c}{\vspace{1mm} \small pansharpened with co-registration}                 \\
\multicolumn{2}{c}{\vspace{2mm} \imCoreg{Q_EL-N5_coreg_1}}                                \\
\vspace{1mm} \small error w/o co-regist. at loss  & \small error with co-regist. at loss  \\
\vspace{1mm} \imCoreg{Q_DT-EL-N5_coreg_1}         & \imCoreg{Q_DT-EL-N5_1}                \\
\end{tabular}
\caption{
Importance of co-registration at loss.
Top: source data. The green/pink out-of-context pixels and strips in the false-color MS are due to band mis-alignment.
Middle: pansharpened image with correct band co-registration.
Bottom: difference (magnified for better visualization) between decimated pansharpened image and MS, $\hM_\downarrow-M$, w/o (left) and with (right) co-registration at loss.
Despite the good visual quality observed in the pansharpened image,
large errors are observed with respect to the MS ({\it e.g.}, pink/green strips near object boundaries) if loss-time band alignment is not performerd.
The corresponding spectral loss will be large, undermining correct model optimization.
}
\label{fig:imgs_coreg}
\end{figure}

For technological reasons, multiresolution images often present some misalignments, also in high-level products.
Different spectral bands are shifted with respect to one another and to the PAN, with errors of up to a few PAN-scale pixels.
Fig.\ref{fig:imgs_coreg} shows on the top an example of this phenomenon:
band misalignment is clearly visible in the MS image, especially near geometric structures, where pixels with unnatural colors appear.
Of course, to obtain a high-quality pansharpened image, such errors must be compensated for.
A possible solution is to co-register the MS bands before proceeding with actual pansharpening.
However, this operation modifies and possibly impairs the original MS reference data.
In addition, it requires the active involvement of the user, not always appropriately skilled.

A better solution is to perform co-registration during pansharpening \cite{Ciotola2022, Seo2020, Lee2021}.
This is accomplished automatically in our full-resolution training framework, with no need for user intervention.
In fact, to minimize the spatial loss, the local correlation between the PAN and each pansharpened band must be large,
and this happens only when the spatial structures in all bands are correctly aligned.
Going back to Fig.\ref{fig:imgs_coreg}, the middle row shows the example image pansharpened by our method.
Misalignment problems have been solved automatically and all bands appear to be correctly co-registered, greatly benefiting visual quality.

However, all our efforts would be undermined if we did not perform the so-called {\em co-registration at loss}.
The example of Fig.\ref{fig:imgs_coreg} will help us describe this issue.
To compute the spectral loss, we must compare the decimated pansharpened image with the original MS.
The former has been automatically co-registered while minimizing the spatial loss, but the latter presents band misalignment.
Therefore, the difference image (bottom-left) will exhibit large errors and large spectral loss.
So, while the spatial loss pushes towards band alignment, the spectral loss discourages it,
a highly undesirable situation in which the two losses work against each other.

Fortunately, once identified, this annoying problem has a simple solution.
We estimate in advance the global shift of each MS band with respect to the reference PAN.
Then, at the moment of computing the spectral loss, and only to this aim, we temporarily shift back the bands of the pansharpened image to align them with the original MS before decimation.
In practice, we reintroduce the band misalignment only to compute a meaningful spectral loss.
The effect is visible in the bottom-right difference image.
Large spectral errors are avoided and spatial and spectral losses are both small, in accordance with the good visual quality of the image.

\subsection{Computing the JESSE loss}

\begin{figure}
\centering
\begin{tabular}{c}
\includegraphics[scale=0.58]{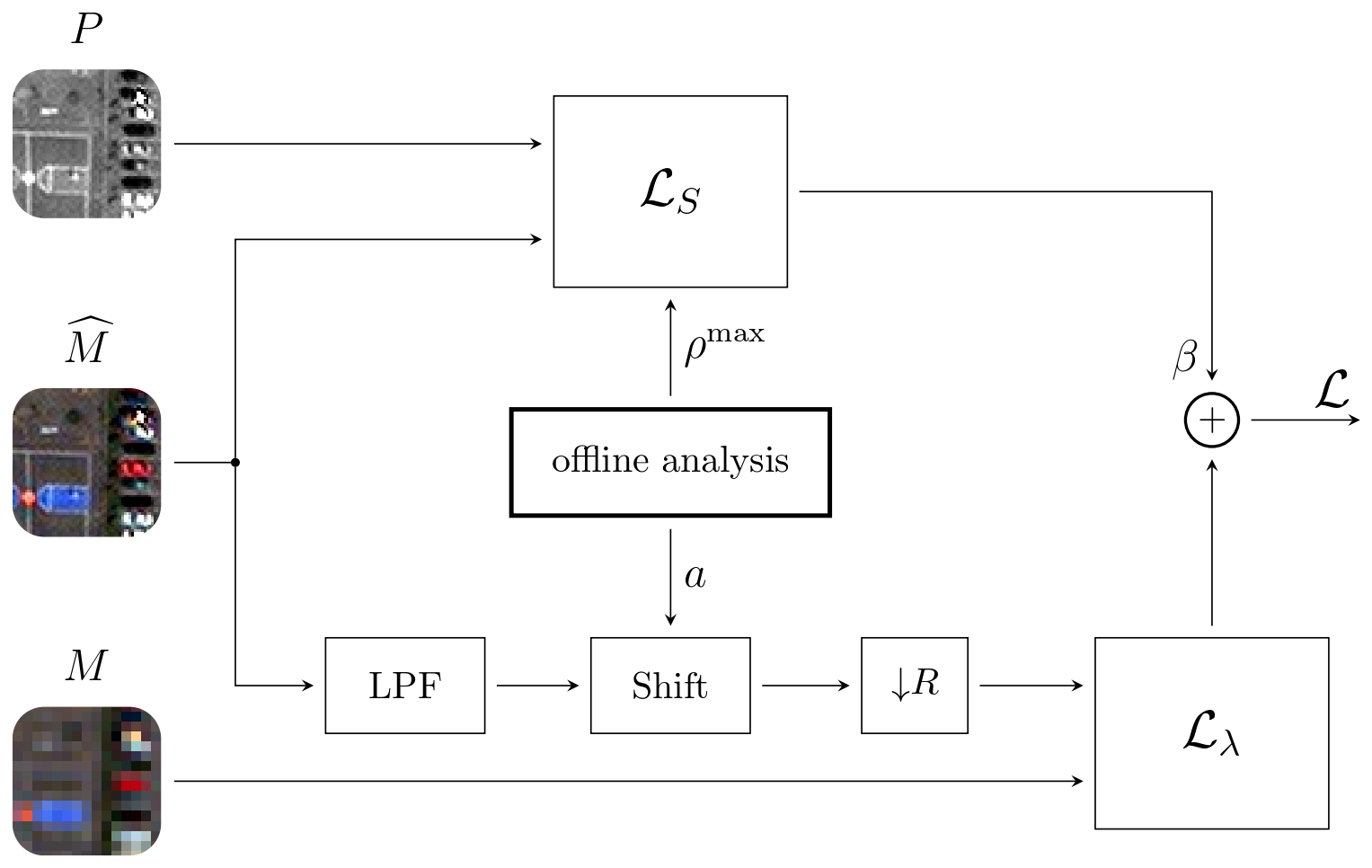}     \\[1mm]
(a) on-line operations                                                      \\[6mm]
\includegraphics[scale=0.58]{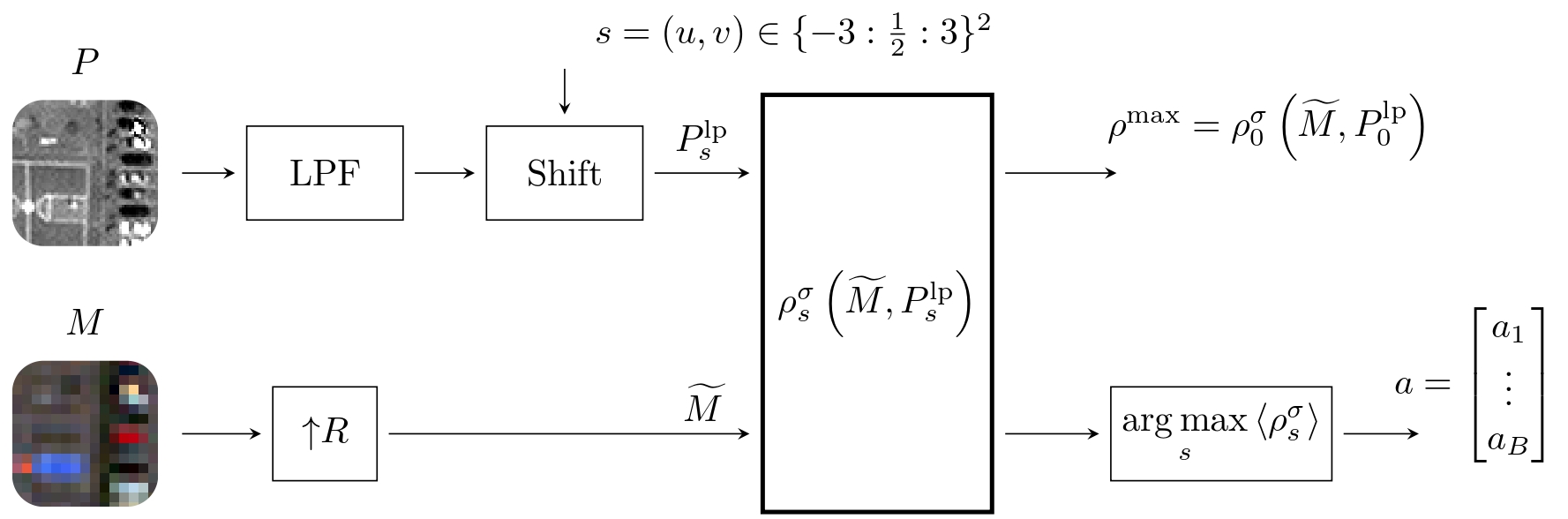} \\[2mm]
(b) off-line analysis
\end{tabular}

\vspace{2mm}
\caption{
Computation of JESSE loss.
(a) on-line operations: the two branches compute spatial and spectral loss components.
(b) off-line correlation-based analysis provides the optimal band shifts and the reference correlation field.
}
\label{fig:coreg_alg}
\end{figure}

We can now summarize the details of the total loss computation with the help of Fig.\ref{fig:coreg_alg}.
The upper part (a) shows a high-level block diagram of on-line operations,
where the top and bottom branches concern spatial and spectral losses, respectively.
Note that both branches rely on products generated in advance through off-line analyses:
the reference correlation map, $\rho^{\max}$, required for the computation of $\LL_S$, and
the vector of band-wise alignment shifts, $a$,         for the computation of $\LL_\lambda$.
This is a qualifying point of this approach.
In fact, since $\rho^{\max}$ and $a$ are functions of $P$ and $M$,
both losses eventually depend, either explicitly or implicitly, on both PAN and MS, that is
\begin{eqnarray}
    \LL_S       & = \LL_S(\hM, P; \rho^{\max})             & = \LL_S^*(\hM, P; M) \\
    \LL_\lambda & = \LL_\lambda(\hM, M; a) \rule{4mm}{0mm} & = \LL_\lambda^*(\hM, M; P)
\end{eqnarray}
Indeed, the correlation and alignment information, computed by exploiting all the available source data,
put in tight relation the spectral and spatial losses, which do not fight with each other anymore
but concur to provide the highest fidelity under all dimensions.
Accordingly, we call this loss JESSE, after {\em Joint Enhancement of Spectral and Spatial fidElity}.

Note also that in the spectral loss defined in eq.(\ref{eq:ll}) $\hM_{\downarrow}$ must be replaced by $\hM_{\downarrow a}$
to account for the band alignment in the computation of the distortion metrics and, accordingly, the total loss becomes

\begin{eqnarray}
    \LL & = & \LL_\lambda + \beta \LL_S = \nonumber \\
        & = & {\DL}(\hM_{\downarrow a}, M) + \gamma {\rm ERGAS}(\hM_{\downarrow a}, M) + \nonumber \\
        &   & + \;\; \beta \left\langle \left(1-\rho^{\sigma}\right) {\rm u}\left(\rho^{\max}-\rho^{\sigma}\right) \right\rangle,
    \label{eq:total}
\end{eqnarray}
with $ \rho^\sigma = {\rm Corr}(P,\hM)$.

Fig.\ref{fig:coreg_alg}(b) describes in detail the off-line correlation-based analysis block.
Since PAN and MS must be compared, they are brought in the same signal space.
The MS is expanded by the resolution ratio $R$, by means of the 23-tap polynomial interpolator proposed in \cite{Aiazzi2002}, to obtain its upscaled version, $\widetilde{M}$.
On the other side, the high-frequency content of $P$, not present in $\widetilde{M}$, is removed through low-pass filtering.
Then, we compute their correlation field using a window of size $R^2$ rather than $R$ to match input images that have been smoothed.
If no shift is applied, we obtain the reference field used to compute the spatial loss.
However, in the presence of band misalignment, the PAN-MS matching may be improved by suitable shifts,
therefore we look for the optimal shift vector $a$ that maximizes the average correlation.
In practice, we perform the search for the optimal shift band-by-band on a discrete grid,
${\cal S}=\{-3:\frac{1}{2}:3\}^2$,
including all displacement with horizontal and vertical components going from $-3$ to $+3$ PAN-scale pixels at half-pixel steps.
In formulas
\begin{equation}
    a_b = \arg\max_{s \in {\cal S}} \langle \rho^{\sigma}_s \rangle = \arg\max_{s \in {\cal S}} \langle {\rm Corr}(P^{lp}_s,\tM(b)) \rangle
    \label{eq:coregatloss}
\end{equation}
with ${\rm Corr}(\cdot,\cdot)$ computed only on valid pixels.

Eventually, the optimal shifts and the reference correlation field are provided as side input for the computation of the JESSE loss.
It is worth underlining that this process does not involve the prediction $\hM$,
hence its computational impact is negligible with respect to the cost of the fine-tuning cycles.

\subsection{Fast Target Adaptation}
\label{sec:TA}

\begin{figure}
\centering
\includegraphics[width=\columnwidth]{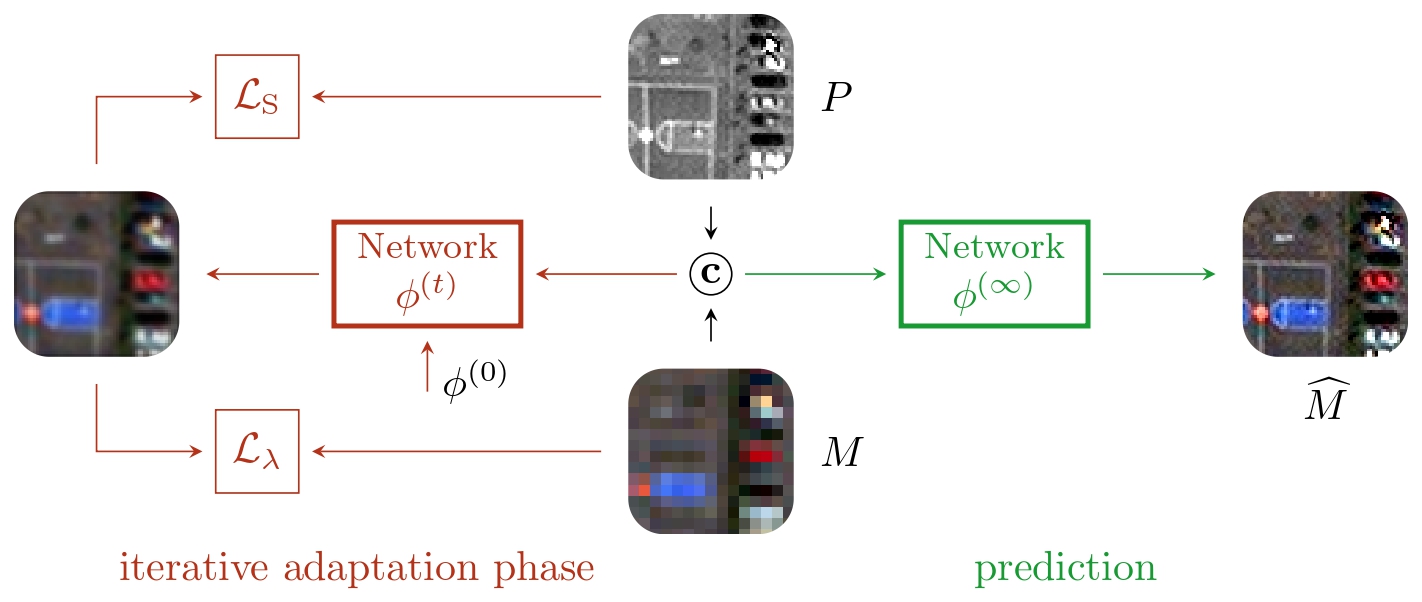} \\
\caption{The full-resolution target-adaptive inference scheme.
The pre-trained weights $\phi^{(0)}$ are iteratively fine-tuned (red blocks) to the target image.
At convergence, the final weights $\phi^{(\infty)}$ are used for actual pansharpening (green block).
}
\label{fig:TA}
\end{figure}

A key asset of the proposed method is the use of a target-adaptive pansharpening modality,
which adapts on-the-fly the pre-trained model to the peculiar statistics of the target image.
This inference-time fine-tuning process was first introduced in \cite{Scarpa2018a} in a supervised context,
and then adapted in \cite{Ciotola2022} for application in a full-resolution unsupervised context.
In Fig.\ref{fig:TA} we show its high-level workflow.
The starting point is a deep learning model with initial weights $\phi^{(0)}$, obtained by off-line pre-training on a suitably large dataset.
However, no matter how large and varied the training set,
these weights will hardly fit the unique features of the target image, in terms of scene, illumination, and acquisition process, including possible band misalignments.
So, to overcome this problem, the model undergoes an iterative tuning process (red blocks on the left of the figure) on the target image itself.
The weights $\phi^{(t)}$ available at time $t$ are used to pansharpen the source image.
Then, the resulting loss is computed (the figure shows separate spectral and spatial terms as customary) and used to update the weights to the next values $\phi^{(t+1)}$.
At the end of the process, the optimized weights, $\phi^{(\infty)}$, are used for actual pansharpening (green block on the right).
Experiments show that target adaptation improves consistently the pre-trained models,
especially when the target image is not well represented by the training set.

In principle,
the adaptation process should proceed until full convergence, but this would significantly delay real-time operations, and only a few tuning steps are carried out in practice.
Moreover, the computational cost grows rapidly with image size and model capacity, preventing its application in important real-world cases.
A first attempt to address this issue \cite{Ciotola2023} provides only minor improvements.
So, we propose here a new adaptation scheme which relies only on a fixed-size tuning sample drawn from the target image,
thereby ensuring a low and almost size-invariant computational complexity.

\begin{figure}
\centering
\includegraphics[width=0.8\columnwidth]{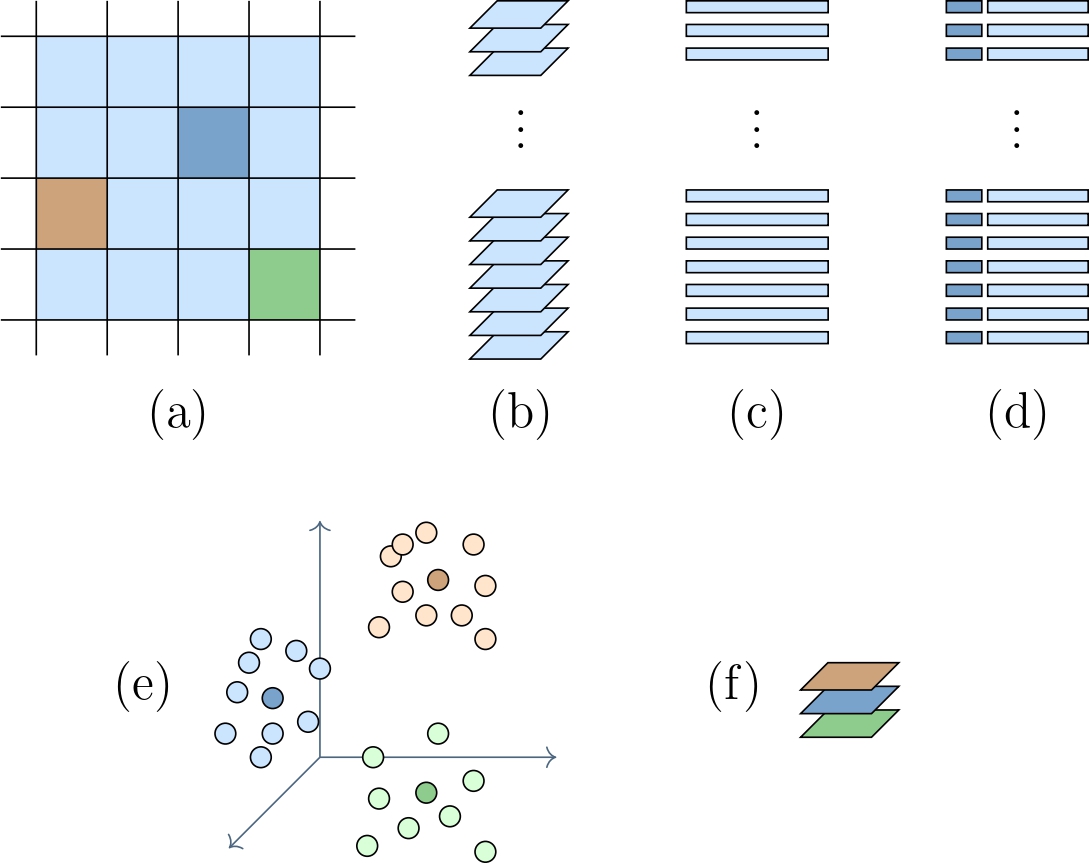} \\
\caption{Selection of the dataset for Fast Target Adaptation:
a) image tiling;
b) initial dataset with all tiles;
c) compact features extracted by the CNN;
d) PCA: the left part is kept;
e) $k$-means clustering and template selection;
f) final dataset comprising only a few tiles representative of all land covers.
}
\label{fig:FastTA}
\end{figure}

\begin{figure*}[!h]
\centering
\newcommand\scal{0.66}
\setlength{\tabcolsep}{6mm}
\begin{tabular}{cc}
\includegraphics[scale = \scal]{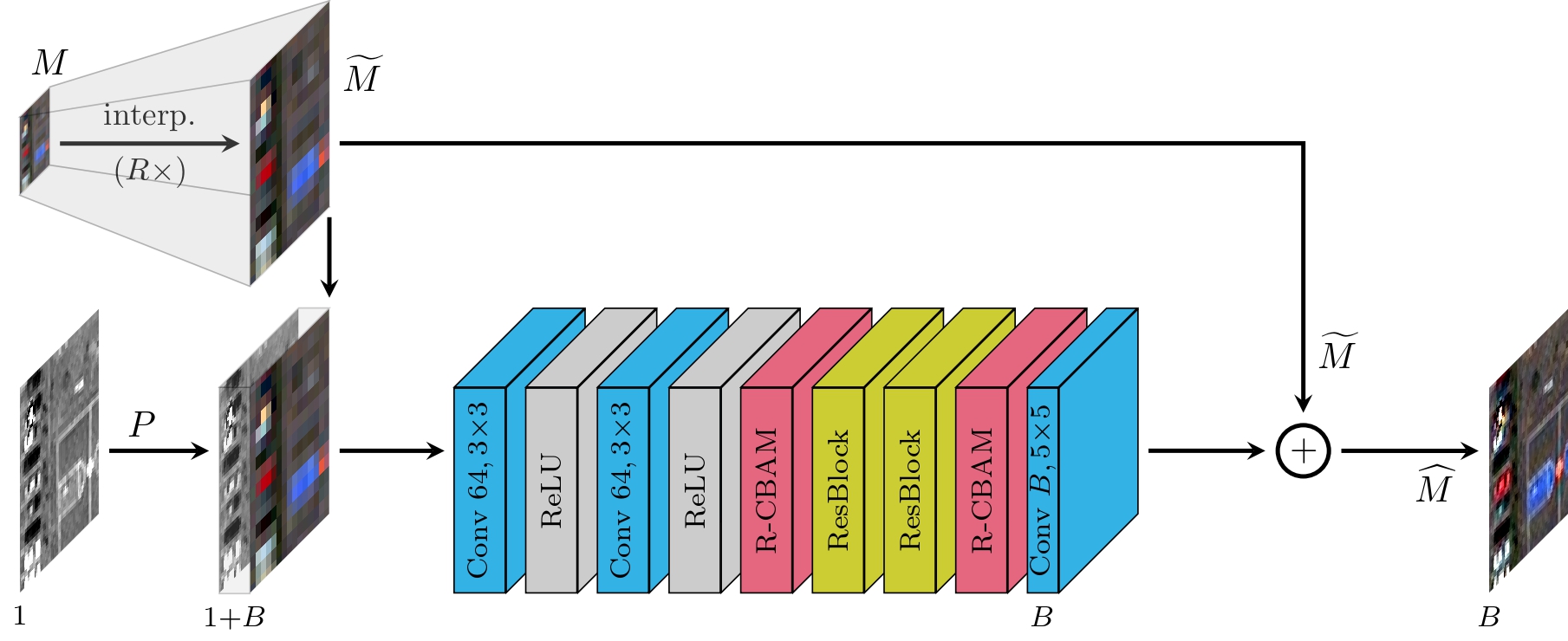} &
\includegraphics[scale = \scal]{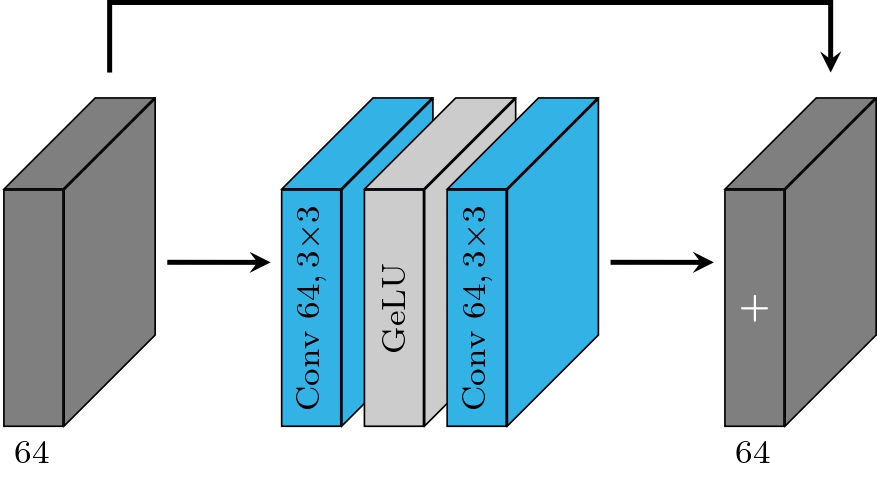} \\
(a) & (b) \\[7mm]
\multicolumn{2}{c}{\includegraphics[scale = \scal]{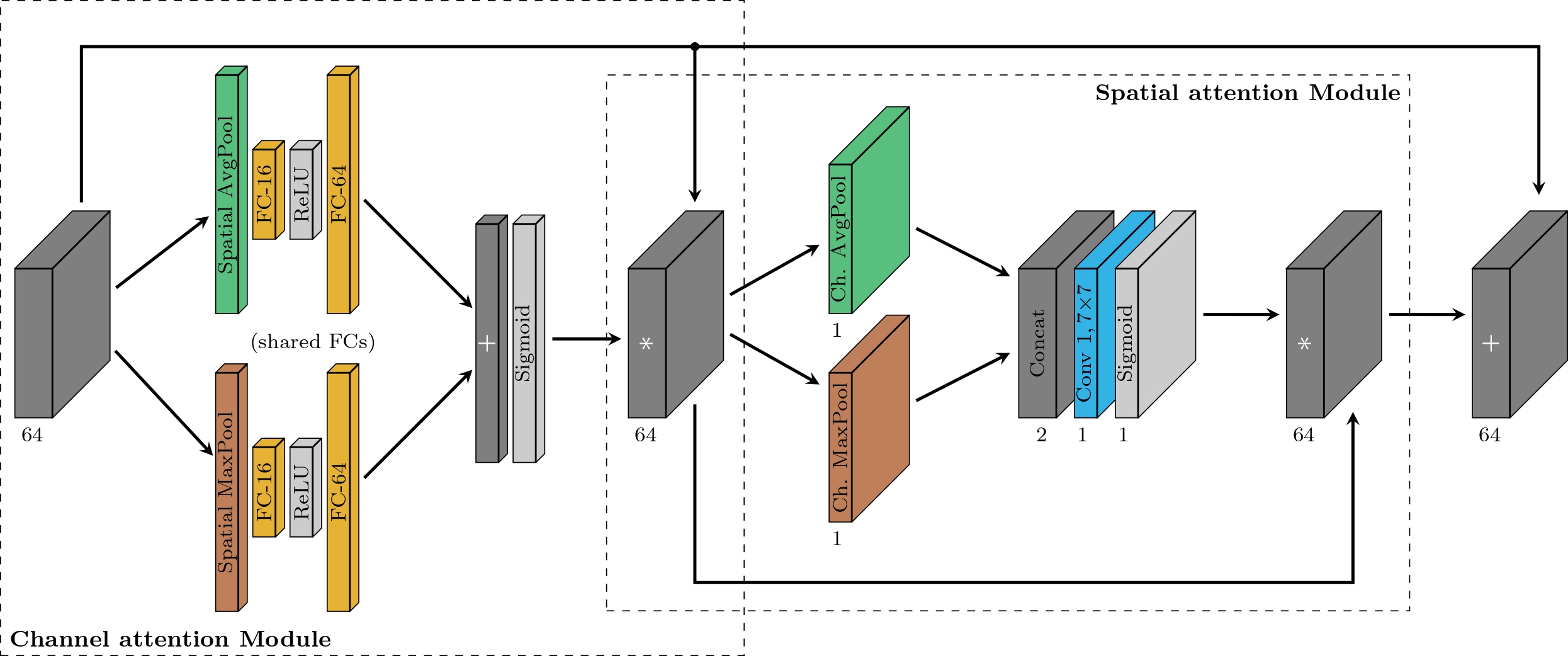}} \\
\multicolumn{2}{c}{(c)}
\end{tabular}
\caption{
The $\lambda$-PNN pansharpening model.
(a) overall architecture;
(b) ResBlock module;
(c) R-CBAM module.}
\label{fig:network}
\end{figure*}

The core of the proposed method is the selection of a suitable fine-tuning dataset,
small enough to allow for fast on-line operations, yet so diverse to capture all the major features of the scene.
This process is described in Fig.\ref{fig:FastTA}.
First, the image (a) is partitioned into $N_{\rm tile}$ tiles of $c{\times}c$-pixels.
Then, the tiles (b) feed a CNN-based classifier, MobileNet v3-small \cite{Howard2019}, which extracts compact 576-component descriptors (c).
These latter are further compacted (d) by means of PCA, retaining only the three principal components.
The resulting vectors, regarded as points in a 3-d space (e), are processed by $k$-means clustering
to obtain $N_{\rm clust} \ll N_{\rm tile}$ groups of similar vectors.
The idea is that each cluster should be representative of a major land cover of the scene.
Therefore, we eventually select a single feature vector per cluster (the median) to represent the whole group
and include the associated tile in the fine-tuning dataset (f).

By acting on tile size and number of clusters, one can make this procedure more or less aggressive.
Experiments described in Section \ref{sec:exp} show that the proposed fast procedure ensures excellent target adaptation
with processing times that are small, and almost invariant to image size.

\subsection{$\lambda$-PNN: a deeper attention-based network architecture}

Thus far,
we have improved the unsupervised training framework, especially from the point of view of spectral quality,
and also the target-adaptive operating modality, which is now more accurate and efficient.

We now turn to propose a new model which takes advantage of the improved unsupervised training framework.
The architecture is depicted in Fig.\ref{fig:network}(a).
Before going into detail, we observe that, with 7 convolutional layers and 2 attention modules, it is significantly deeper than our previous proposals.
This is a direct consequence of the fast target-adaptive modality described above,
which allows us to fine-tune a much heavier network than before without any major impact on the processing time.

Our architecture is of the residual type \cite{He2016}:
a global skip connection brings the resized MS directly to the output,
where it is added to the output of the convolutional branch.
In addition, also most of the convolutional blocks have a residual structure.
This choice reflects the intuition that part of the desired image is already available (the low-frequency MS component),
and only the high-frequency detail should be actually predicted.
On the other hand, most deep learning models proposed for pansharpening in latest years \cite{Yang2017, Wei2017L, Yuan2018, Deng2020, Luo2020, Ciotola2022},
rely on residual architectures, with advantages in terms of training speed and generalization properties.

In the convolutional branch, there are two 64-channel convolutional layers with ReLU activations,
followed by two residual blocks Fig.\ref{fig:network}(b),
comprising again two 64-channel convolutional layers each and GeLU activation \cite{Hendrycks2016},
and a final $B$-channel convolutional layer with linear activation feeding the summing node.
In addition, there are two convolutional block attention modules (CBAM) \cite{Woo2018}, these too in residual configuration.
The R-CBAM modules aim of focusing attention on especially relevant portions of the input, both in space and along the channels.
Their architecture is depicted in Fig.\ref{fig:network}(c).
In the channel attention module,
global spatial pooling (max and average) is performed on two parallel paths followed by a shared multilayer perceptron.
The resulting vectors, summed to one another and squashed on the 0-1 interval by a sigmoid,
encode the channel importance.
Accordingly, they multiply the 64-channel input stack emphasizing some channels more than others.
The spatial attention module performs in a similar way, by exchanging the role of space and channel.
After channel-wise max and average pooling
the resulting feature maps are compacted to a single heatmap by a convolutional layer with sigmoid activation.
This spatial attention map then multiplies the channel-emphasized input to put emphasis also on selected spatial sites.
To help prevent vanishing gradients, the whole CBAM has a residual architecture,
with the resulting feature stack summed to the input to provide the final output.

\section{Experimental Analysis}
\label{sec:exp}

In this Section,
after analyzing a sample experiment that provides more insight into the JESSE loss,
we carry out the comparative performance assessment of the proposed method, studying numerical and visual results.
Then, we show ablation studies that validate our design choices,
and assess the efficiency of the new fast target adaptation procedure.

\subsection{Datasets}

\begin{table}
\centering
\setlength{\tabcolsep}{4pt}
\begin{tabular}{lcccc}
\multicolumn{5}{c}{{\bf WorldView-3} (GSD at nadir: 0.31m)}                                     \\ \hline
Dataset           &     Training     &        Val.      &       X-Val.     &        Test        \\
(PAN size)        & (512$\times$512) & (512$\times$512) & (512$\times$512) & (2048$\times$2048) \\ \hline
Fortaleza         &               32 &                8 &                - &                  - \\
Mexico City       &               32 &                8 &                - &                  - \\
Xian              &               32 &                8 &                - &                  - \\
Adelaide          &                - &                - &               24 &                  8 \\
Munich (PairMax)  &                - &                - &                - &                  3 \\ \hline \\
\multicolumn{5}{c}{{\bf WorldView-2} (GSD at nadir: 0.46m)}                                     \\ \hline
Dataset           &     Training     &        Val.      &       X-Val.     &        Test        \\
(PAN size)        & (512$\times$512) & (512$\times$512) & (512$\times$512) & (2048$\times$2048) \\ \hline
Berlin            &               32 &                8 &                - &                  - \\
London            &               32 &                8 &                - &                  - \\
Rome              &               32 &                8 &                - &                  - \\
Washington        &                - &                - &               24 &                 10 \\
Miami (PairMax)   &                - &                - &                - &                  2 \\ \hline \\
\multicolumn{5}{c}{{\bf GeoEye-1} (GSD at nadir: 0.41m)}                                        \\ \hline
Dataset           &     Training     &        Val.      &       X-Val.     &        Test        \\
(PAN size)        & (512$\times$512) & (512$\times$512) & (512$\times$512) & (2048$\times$2048) \\ \hline
Norimberga        &               32 &                8 &                - &                  - \\
Rome              &               32 &                8 &                - &                  - \\
Waterford         &               32 &                8 &                - &                  - \\
Genoa             &                - &                - &               24 &                  9 \\
London (PairMax)  &                - &                - &                - &                  1 \\
Trenton (PairMax) &                - &                - &                - &                  1 \\ \hline
\end{tabular}
\caption{
WV3, WV2, and GE1 datasets, with number of crops for training, validation and test.
Adelaide and Washington, courtesy of DigitalGlobe$^\copyright$.
Fortaleza, Mexico City, Xian, Berlin, London (WV2), Rome, Norimberga, Waterford and Genoa (DigitalGlobe$^\copyright$) provided by ESA.
Munich, Miami, London (GE1) and Trenton are part of PairMax dataset \cite{Vivone2021}.
}
\label{tab:datasets}
\end{table}

To obtain a reliable wide-spectrum assessment, we carry out our experiments on a large variety of data,
making also sure to test the generalization ability of competing methods.
Therefore, we consider three distinct datasets, one for each of the sensors WorldView-3, WorldView-2, and GeoEye-1.
For each dataset,
we have several large images available (see Tab.\ref{tab:datasets}).
Three of them are used exclusively for training and validation.
Another image is never seen in training, but used to gather ``cross-scenario'' (change of place, date, daylight conditions, sensing geometry)
validation information and then also for testing.
A last image is used exclusively for testing.
This latter is the most challenging case since there is no link between the test image and the training process.
Training and validation are carried out on 512$\times$512-pixel crops (PAN resolution)
while there is no size constraint in the testing phase and we consider 2048$\times$2048 crops.
For all sensors, the PAN-MS resolution ratio is $R=4$.
In the following, we name datasets after the corresponding sensor: WV3, WV2, GE1, with suffix Train, Val, X-Val, and Test when appropriate.

\subsection{Reference Methods}

\begin{table}
\centering
\setlength{\tabcolsep}{4pt}
\begin{tabular}{p{8.4cm}}
\multicolumn{1}{l}{\ru \bf Component Substitution (CS)}              \\ \hline \ru
BT-H \cite{Lolli2017}, BDSD \cite{Garzelli2008}, BDSD-PC \cite{Vivone2019}, GS \cite{Laben2000}, GSA \cite{Aiazzi2007}, PRACS \cite{Choi2011} \vspace{2mm} \\
\multicolumn{1}{l}{\ru \bf Multiresolution Analysis (MRA)}           \\ \hline \ru
AWLP \cite{Otazu2005}, MTF-GLP \cite{Alparone2017}, MTF--FS \cite{Vivone2018a}, MTF--HPM \cite{Alparone2017}, MF \cite{Restaino2016} \vspace{2mm} \\
\multicolumn{1}{l}{\ru \bf Variational Optimization (VO)}            \\ \hline \ru
FE-HPM \cite{Vivone2015a}, SR-D \cite{Vicinanza2015}, TV \cite{Palsson2014} \vspace{2mm} \\
\multicolumn{1}{l}{\ru \bf Machine Learning (ML) Reduced Resolution} \\ \hline \ru
PNN \cite{Masi2016}, A-PNN \cite{Scarpa2018a}, A-PNN-TA \cite{Scarpa2018a}, BDPN \cite{Zhan2019}, DiCNN \cite{He2019}, DRPNN \cite{Wei2017L}, FusionNet \cite{Deng2020}, MSDCNN \cite{Yuan2018}, PanNet \cite{Yang2017} \vspace{2mm} \\
\multicolumn{1}{l}{\ru \bf Machine Learning (ML) Full Resolution}    \\ \hline \ru
SSQ \cite{Luo2020}, GDD \cite{Uezato2020}, PanGan \cite{Ma2020}, Z-PNN \cite{Ciotola2022} \\
\end{tabular}
\caption{Reference methods used for comparative analysis.}
\label{tab:methods}
\end{table}

We compare the performance of the proposed $\lambda$-PNN with a large number of reference methods.
They are listed in Tab.\ref{tab:methods}, grouped by their general approach:
component substitution, multiresolution analysis, variational optimization, machine learning,
the latter trained at reduced or full resolution.
Most of the methods are available in the toolboxes \cite{Vivone2020} and \cite{Deng2022}, from which we selected those that performed best in the experiments.
Methods of the last group, instead, were downloaded from the authors' websites.
For machine learning methods,
together with the source code, the authors usually provide the weights obtained on their own training sets.
In some cases (marked by an asterisk in the tables of results), weights were not available and we re-trained the models on our dataset.
Finally, we re-implemented and trained SSQ (marked with a double asterisk) because neither code nor weights were available.

\subsection{Performance Metrics}

In pansharpening, the lack of ground truths does not impact only model training but also performance assessment.
All measures of distortion are necessarily indirect and rely on explicit or implicit hypotheses that remain to be proved.
Based on our reasoning and experience,
we believe that the metrics used in $\lambda$-PNN to compute the total loss, that is, $\DLa$, R-ERGAS, and $\DR=\langle(1-\rho^{\sigma}) {\rm u}(\rho^{\max}-\rho^{\sigma})\rangle$,
provide a meaningful assessment of spectral and spatial fidelity.
However, for the sake of completeness,
we consider some more metrics largely used in the literature:
Khan's measure of spectral distortion without band alignment, $\DL$,
and two more spatial distortion metrics, $\DS$ \cite{Alparone2008} and $\DSR$ \cite{Alparone2018}, for which description we refer the reader to the original papers.

\subsection{On the coherency of spatial and spectral losses}

\begin{figure}
\centering
\includegraphics[scale=0.84]{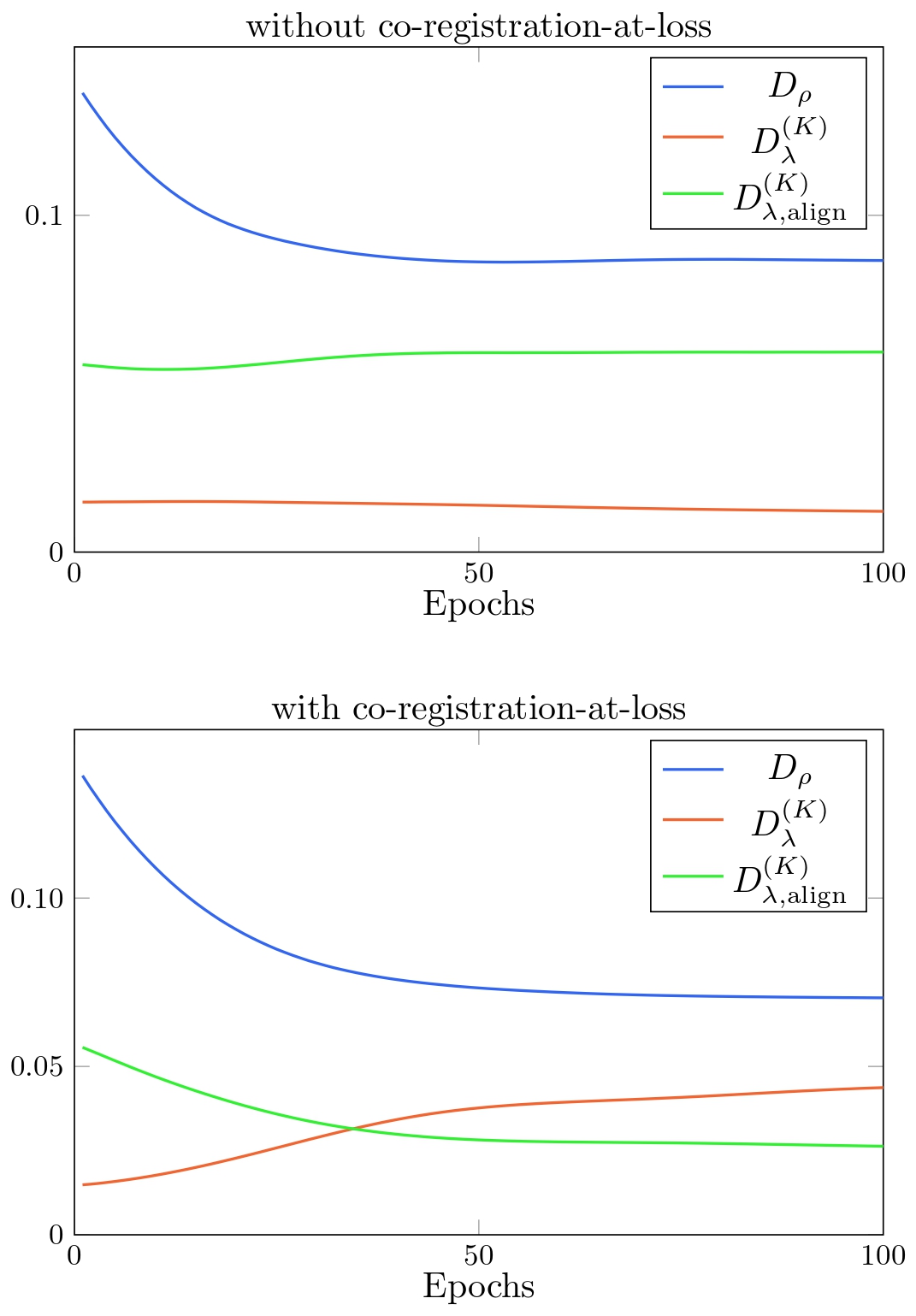}
\caption{
Evolution of distortion metrics: $\DR$ (blue), $\DL$ (red), $\DLa$ (green), with target adaptation.
Without re-alignment (top), target adaptation is almost useless: $\DLa$ is always very high, indicating a large spectral distortion.
With re-alignment (bottom), it becomes very effective: $\DLa$ reduces significantly and also $\DR$ gets smaller than before.
}
\label{fig:align_no_yes}
\end{figure}

To gain some more insight on the proposed JESSE loss, let us pansharpen with $\lambda$-PNN the image shown in Fig.\ref{fig:imgs_coreg},
affected by obvious alignment problems.
The network has been already pre-trained on a large dataset and now must be fine-tuned on the target image.
In Fig.\ref{fig:align_no_yes} we show the evolution of spatial ($\DR$) and spectral ($\DLa$) distortion\footnote{ERGAS is not shown to avoid cluttering the plots.}
as the adaptation proceeds,

In the top plot, however, we use a ``wrong'' loss for adaptation, with the alignment vector set to $a=0$,
such that the pansharpened image is compared with the original MS without compensating the shifts.
In other words, we replace $\DLa$ with $\DL$ in the loss.
The effect of this wrong choice is to render target adaptation almost ineffective.
First of all, $\DLa$ (green curve) does not improve at all, it even worsens a little bit,
because the network is trying instead to further improve $\DL$ (red), which was already optimized off-line.
In addition,
the spatial distortion $\DR$ (blue) plateaus at 0.086 after a few cycles,
because increasing the correlation with the PAN would align the bands and then increase the distortion with respect to the original MS, hence $\DL$.
In short, spatial and spectral loss components are fighting against each other.

In the bottom plot, instead, the correct loss is used, with the optimal alignment vector $a$.
Now, as the adaptation proceeds, the true measure of spectral distortion, $\DLa$, reduces constantly, going from 0.58 to 0.25,
while $\DL$ grows.
In addition, also the spatial distortion, $\DR$, keeps lowering, reaching 0.071 after 100 cycles.
Spectral and spatial losses provide coherent indications and reinforce one another, concurring to the joint enhancement of spectral and spatial fidelity.

\subsection{Numerical Results}

\newcommand{\za}[1]{\textbf{#1}}
\newcommand{\zb}[1]{\underline{#1}}
\newcommand{\zr}{\rule{0mm}{3.5mm}}

\begin{table*}
\footnotesize
\centering
\setlength{\tabcolsep}{3pt}
\begin{tabular}{lc@{\rule{6mm}{0mm}}ccccccc@{\rule{6mm}{0mm}}cccccc} \hline
 \zr              &   &     \multicolumn{6}{c}{Adelaide}                                                  &   &     \multicolumn{6}{c}{Munich (PairMax)}                                          \\             \cline{3-8} \cline{10-15}
 \zr Method       &   &     $\DLa$  &    \RERGAS  &   ~~$\DL$~~ &   ~~$\DR$~~ &   ~~$\DS$~~ & ~~$\DSR$~~  &   &     $\DLa$  &    \RERGAS  &   ~~$\DL$~~ &   ~~$\DR$~~ &   ~~$\DS$~~ &  ~~$\DSR$~~ \\ \cline{1-1} \cline{3-8} \cline{10-15}
 \zr EXP          &   &     0.069   &     3.853   &     0.049   &     0.859   &     0.114   &     0.179   &   &     0.064   &     3.911   &     0.075   &     0.854   &     0.132   &     0.161   \\
 BT-H             &   &     0.074   &     3.685   &     0.153   &     0.071   &     0.123   & \za{0.001}  &   &     0.079   &     4.161   &     0.156   & \zb{0.049}  &     0.122   & \za{0.001}  \\
 BDSD             &   &     0.100   &     4.392   &     0.175   &     0.110   &     0.047   &     0.016   &   &     0.120   &     5.596   &     0.204   &     0.075   &     0.049   &     0.052   \\
 BDSD-PC          &   &     0.078   &     3.929   &     0.159   &     0.075   &     0.068   &     0.012   &   &     0.108   &     5.447   &     0.190   &     0.058   &     0.057   &     0.048   \\
 GS               &   &     0.113   &     4.370   &     0.199   &     0.086   &     0.085   & \zb{0.003}  &   &     0.128   &     5.012   &     0.213   &     0.072   &     0.083   & \zb{0.001}  \\
 GSA              &   &     0.066   &     3.698   &     0.142   & \zb{0.069}  &     0.112   &     0.004   &   &     0.075   &     4.416   &     0.164   &     0.051   &     0.093   &     0.001   \\
 PRACS            &   &     0.061   &     3.546   &     0.080   &     0.191   &     0.051   &     0.030   &   &     0.063   &     3.919   &     0.099   &     0.195   &     0.053   &     0.023   \\
 AWLP             &   &     0.049   &     3.237   & \zb{0.038}  &     0.093   &     0.069   &     0.059   &   &     0.043   &     3.078   & \zb{0.053}  &     0.079   &     0.076   &     0.039   \\
 MTF-GLP          &   &     0.049   &     3.259   &     0.043   &     0.073   &     0.100   &     0.055   &   &     0.042   &     2.948   &     0.056   &     0.056   &     0.098   &     0.035   \\
 MTF-GLP-FS       &   &     0.051   &     3.311   &     0.041   &     0.097   &     0.088   &     0.054   &   &     0.043   &     2.979   &     0.056   &     0.067   &     0.088   &     0.034   \\
 MTF-GLP-HPM      &   &     0.052   &     3.330   &     0.054   &     0.081   &     0.088   &     0.059   &   &     0.048   &     3.383   &     0.069   &     0.061   &     0.092   &     0.039   \\
 MF               &   &     0.046   &     3.094   &     0.058   &     0.093   &     0.093   &     0.058   &   &     0.042   &     3.037   &     0.065   &     0.078   &     0.085   &     0.051   \\
 FE-HPM           &   &     0.050   &     3.320   &     0.046   &     0.093   &     0.087   &     0.062   &   &     0.043   &     3.103   &     0.057   &     0.073   &     0.089   &     0.043   \\
 SR-D             &   &     0.054   &     3.542   & \za{0.023}  &     0.301   &     0.032   &     0.133   &   & \zb{0.034}  & \zb{2.777}  & \za{0.040}  &     0.186   &     0.070   &     0.086   \\
 TV               &   & \zb{0.036}  & \zb{2.646}  &     0.058   &     0.205   &     0.035   &     0.049   &   &     0.040   &     2.888   &     0.074   &     0.168   &     0.079   &     0.033   \\   \cline{1-1} \cline{3-8} \cline{10-15}
 \zr PNN          &   &     0.197   &     6.675   &     0.269   &     0.461   &     0.073   &     0.116   &   &     0.416   &     9.115   &     0.548   &     0.475   &     0.105   &     0.118   \\
 A-PNN            &   &     0.070   &     3.591   &     0.100   &     0.534   &     0.082   &     0.096   &   &     0.169   &     4.398   &     0.274   &     0.665   &     0.187   &     0.167   \\
 A-PNN-FT         &   &     0.060   &     3.489   &     0.068   &     0.332   & \zb{0.026}  &     0.069   &   &     0.085   &     3.521   &     0.121   &     0.335   & \za{0.029}  &     0.084   \\
 BDPN             &   &     0.150   &     5.393   &     0.261   &     0.178   &     0.078   &     0.008   &   &     0.294   &     7.693   &     0.440   &     0.311   &     0.115   &     0.024   \\
 DiCNN            &   &     0.158   &     5.602   &     0.246   &     0.414   &     0.082   &     0.056   &   &     0.217   &     6.077   &     0.291   &     0.410   &     0.092   &     0.080   \\
 DRPNN            &   &     0.151   &     5.230   &     0.243   &     0.195   &     0.067   &     0.011   &   &     0.198   &     6.598   &     0.304   &     0.186   &     0.086   &     0.012   \\
 FusionNet        &   &     0.099   &     4.228   &     0.151   &     0.412   &     0.067   &     0.094   &   &     0.238   &     5.302   &     0.328   &     0.340   &     0.058   &     0.125   \\
 MSDCNN           &   &     0.157   &     5.450   &     0.243   &     0.226   &     0.098   &     0.013   &   &     0.392   &     6.664   &     0.514   &     0.348   &     0.098   &     0.023   \\
 PanNet           &   &     0.055   &     3.436   &     0.045   &     0.338   & \za{0.014}  &     0.087   &   &     0.061   &     3.322   &     0.082   &     0.297   & \zb{0.035}  &     0.061   \\
 SSQ**            &   &     0.060   &     3.539   &     0.059   &     0.271   &     0.045   &     0.028   &   &     0.083   &     3.567   &     0.118   &     0.284   &     0.073   &     0.018   \\
 GDD*             &   &     0.310   &     9.731   &     0.377   &     0.662   &     0.111   &     0.118   &   &     0.304   &     8.693   &     0.400   &     0.587   &     0.095   &     0.104   \\
 PanGan*          &   &     0.143   &     4.825   &     0.284   &     0.130   &     0.079   &     0.048   &   &     0.630   &     17.589  &     0.773   &     0.132   &     0.117   &     0.070   \\
 Z-PNN            &   &     0.044   &     2.834   &     0.106   &     0.088   &     0.119   &     0.032   &   &     0.091   &     3.756   &     0.153   &     0.101   &     0.141   &     0.033   \\   \cline{1-1} \cline{3-8} \cline{10-15}
 \zr \LPNN        &   & \za{0.021}  & \za{1.978}  &     0.095   & \za{0.044}  &     0.083   &     0.076   &   & \za{0.031}  & \za{2.526}  &     0.066   & \za{0.033}  &     0.140   &     0.079   \\   \hline
\end{tabular}
\caption{Average results on WV3-Test. Left: Adelaide. Right: Munich (PairMax)}
\label{tab:WV3 results}
\end{table*}

\begin{table*}
\footnotesize
\centering
\setlength{\tabcolsep}{3pt}
\begin{tabular}{lc@{\rule{6mm}{0mm}}ccccccc@{\rule{6mm}{0mm}}cccccc} \hline
 \zr              &   &    \multicolumn{6}{c}{Washington}                                                 &   &  \multicolumn{6}{c}{Miami (PairMax)}                                        \\             \cline{3-8} \cline{10-15}
 \zr Method       &   &     $\DLa$  &    \RERGAS  &   ~~$\DL$~~ &   ~~$\DR$~~ &   ~~$\DS$~~ &  ~~$\DSR$~~ &   &     $\DLa$  &    \RERGAS  &   ~~$\DL$~~ &   ~~$\DR$~~ &   ~~$\DS$~~ &  ~~$\DSR$~~ \\ \cline{1-1} \cline{3-8} \cline{10-15}
 \zr EXP          &   &     0.042   &     2.238   &     0.044   &     0.808   &     0.069   &     0.151   &   &     0.056   &     3.663   &     0.057   &     0.738   &     0.065   &     0.135   \\
 BT-H             &   &     0.056   &     2.428   &     0.101   &     0.071   &     0.114   & \za{0.000}  &   &     0.064   &     3.819   &     0.111   &     0.110   &     0.087   & \za{0.000}  \\
 BDSD             &   &     0.208   &     4.679   &     0.325   &     0.186   &     0.093   &     0.091   &   &     0.118   &     5.303   &     0.201   &     0.237   &     0.070   &     0.049   \\
 BDSD-PC          &   &     0.107   &     3.335   &     0.181   &     0.088   &     0.040   &     0.055   &   &     0.095   &     4.742   &     0.161   &     0.138   &     0.030   &     0.046   \\
 GS               &   &     0.108   &     3.378   &     0.185   &     0.110   &     0.096   &     0.001   &   &     0.086   &     4.503   &     0.142   &     0.149   &     0.089   & \zb{0.001}  \\
 GSA              &   &     0.058   &     2.505   &     0.108   &     0.081   &     0.113   & \zb{0.001}  &   &     0.063   &     3.946   &     0.114   &     0.108   &     0.077   &     0.002   \\
 PRACS            &   &     0.042   &     2.214   &     0.056   &     0.279   &     0.034   &     0.052   &   &     0.056   &     3.620   &     0.074   &     0.229   &     0.036   &     0.017   \\
 AWLP             &   &     0.029   &     1.829   & \zb{0.033}  &     0.086   &     0.079   &     0.045   &   &     0.033   &     2.870   & \zb{0.036}  &     0.129   &     0.059   &     0.061   \\
 MTF-GLP          &   &     0.031   &     1.836   &     0.043   &     0.065   &     0.109   &     0.045   &   &     0.033   &     2.841   &     0.039   &     0.105   &     0.077   &     0.056   \\
 MTF-GLP-FS       &   &     0.031   &     1.853   &     0.038   &     0.098   &     0.099   &     0.038   &   &     0.035   &     2.909   &     0.039   &     0.129   &     0.069   &     0.053   \\
 MTF-GLP-HPM      &   &     0.057   &     2.375   &     0.071   &     0.076   &     0.094   &     0.047   &   &     0.038   &     2.951   &     0.058   &     0.129   &     0.060   &     0.060   \\
 MF               &   &     0.039   &     1.958   &     0.062   &     0.083   &     0.103   &     0.052   &   &     0.032   &     2.817   &     0.045   &     0.130   &     0.069   &     0.066   \\
 FE-HPM           &   &     0.040   &     2.068   &     0.049   &     0.082   &     0.100   &     0.050   &   &     0.033   &     2.893   &     0.042   &     0.122   &     0.073   &     0.058   \\
 SR-D             &   & \zb{0.025}  &     1.753   & \za{0.018}  &     0.245   &     0.031   &     0.083   &   &     0.028   &     2.719   & \za{0.027}  &     0.258   &     0.041   &     0.123   \\
 TV               &   &     0.047   & \zb{1.616}  &     0.086   &     0.241   & \za{0.025}  &     0.042   &   & \za{0.016}  & \za{2.063}  &     0.039   &     0.274   & \za{0.023}  &     0.051   \\   \cline{1-1} \cline{3-8} \cline{10-15}
 \zr PNN          &   &     0.073   &     2.349   &     0.110   &     0.249   &     0.029   &     0.011   &   &     0.055   &     3.575   &     0.083   &     0.335   &     0.027   &     0.015   \\
 A-PNN            &   &     0.047   &     2.097   &     0.072   &     0.419   &     0.040   &     0.036   &   &     0.044   &     3.302   &     0.058   &     0.508   &     0.052   &     0.046   \\
 A-PNN-FT         &   &     0.043   &     2.076   &     0.064   &     0.274   & \zb{0.025}  &     0.021   &   &     0.040   &     3.177   &     0.053   &     0.338   &     0.026   &     0.027   \\
 BDPN             &   &     0.093   &     2.649   &     0.150   &     0.195   &     0.047   &     0.013   &   &     0.079   &     4.548   &     0.140   &     0.215   &     0.046   &     0.013   \\
 DiCNN            &   &     0.067   &     2.492   &     0.115   &     0.302   &     0.045   &     0.026   &   &     0.091   &     5.258   &     0.166   &     0.413   &     0.042   &     0.051   \\
 DRPNN            &   &     0.063   &     2.258   &     0.105   &     0.233   &     0.035   &     0.013   &   &     0.054   &     3.512   &     0.091   &     0.279   & \zb{0.024}  &     0.016   \\
 FusionNet        &   &     0.057   &     2.170   &     0.091   &     0.305   &     0.039   &     0.028   &   &     0.046   &     3.194   &     0.080   &     0.367   &     0.028   &     0.054   \\
 MSDCNN           &   &     0.082   &     2.429   &     0.138   &     0.231   &     0.047   &     0.010   &   &     0.060   &     3.607   &     0.106   &     0.253   &     0.031   &     0.023   \\
 PanNet           &   &     0.041   &     1.839   &     0.060   &     0.345   &     0.031   &     0.046   &   &     0.030   &     2.704   &     0.042   &     0.368   &     0.036   &     0.072   \\
 SSQ**            &   &     0.043   &     1.938   &     0.064   &     0.302   &     0.035   &     0.026   &   &     0.033   &     2.883   &     0.043   &     0.386   &     0.030   &     0.038   \\
 GDD*             &   &     0.282   &     6.351   &     0.369   &     0.762   &     0.080   &     0.124   &   &     0.228   &     7.640   &     0.317   &     0.449   &     0.093   &     0.099   \\
 PanGan*          &   &     0.244   &     5.616   &     0.341   &     0.203   &     0.059   &     0.106   &   &     0.165   &     6.513   &     0.247   &     0.207   &     0.070   &     0.069   \\
 Z-PNN            &   &     0.047   &     1.924   &     0.094   & \zb{0.046}  &     0.119   &     0.021   &   &     0.050   &     3.155   &     0.095   & \zb{0.080}  &     0.095   &     0.026   \\   \cline{1-1} \cline{3-8} \cline{10-15}
 \zr \LPNN        &   & \za{0.020}  & \za{1.291}  &     0.051   & \za{0.042}  &     0.094   &     0.058   &   & \zb{0.024}  & \zb{2.246}  &     0.055   & \za{0.050}  &     0.068   &     0.086   \\   \hline
\end{tabular}
\caption{Average results on WV2-Test. Left: Washington. Right: Miami (PairMax)}
\label{tab:WV2 results}
\end{table*}

\begin{table*}
\footnotesize
\centering
\setlength{\tabcolsep}{3pt}
\begin{tabular}{lc@{\rule{6mm}{0mm}}ccccccc@{\rule{6mm}{0mm}}cccccc} \hline
 \zr              &   &  \multicolumn{6}{c}{Genoa}                                                        &   &  \multicolumn{6}{c}{London+Trenton (PairMax)}                               \\             \cline{3-8} \cline{10-15}
 \zr Method       &   &      $\DLa$  &    \RERGAS &   ~~$\DL$~~ &  ~~$\DR$~~  &  ~~$\DS$~~  & ~~$\DSR$~~  &   &     $\DLa$  &  \RERGAS    &   ~~$\DL$~~ &   ~~$\DR$~~ &   ~~$\DS$~~ &  ~~$\DSR$~~ \\ \cline{1-1} \cline{3-8} \cline{10-15}
 \zr EXP          &   &     0.136    &     4.463  &     0.120   &     0.805   &     0.096   &     0.225   &   &     0.063   &     4.434   &     0.093   &     0.805   &     0.078   &     0.133   \\
 BT-H             &   &     0.158    &     4.579  &     0.261   &     0.070   &     0.091   &     0.012   &   &     0.100   &     5.240   &     0.179   &     0.057   &     0.080   & \zb{0.004}  \\
 BDSD             &   &     0.166    &     4.497  &     0.282   &     0.082   &     0.078   &     0.015   &   &     0.137   &     6.230   &     0.231   &     0.065   &     0.044   &     0.038   \\
 BDSD-PC          &   &     0.167    &     4.495  &     0.284   &     0.081   &     0.079   &     0.015   &   &     0.136   &     6.193   &     0.229   &     0.062   &     0.048   &     0.037   \\
 GS               &   &     0.186    &     4.615  &     0.296   &     0.092   &     0.088   & \za{0.002}  &   &     0.108   &     5.543   &     0.183   &     0.057   &     0.074   & \za{0.002}  \\
 GSA              &   &     0.161    &     4.446  &     0.263   & \za{0.052}  &     0.152   & \zb{0.008}  &   &     0.104   &     5.306   &     0.185   & \za{0.039}  &     0.106   &     0.002   \\
 PRACS            &   &     0.126    &     4.199  &     0.147   &     0.246   & \za{0.053}  &     0.087   &   &     0.072   &     4.673   &     0.125   &     0.183   &     0.062   &     0.027   \\
 AWLP             &   &     0.098    &     3.816  & \zb{0.088}  &     0.082   &     0.110   &     0.108   &   &     0.041   &     3.704   &     0.059   &     0.098   &     0.058   &     0.051   \\
 MTF-GLP          &   &     0.092    &     3.696  &     0.092   &     0.065   &     0.130   &     0.109   &   &     0.037   &     3.498   &     0.055   &     0.078   &     0.066   &     0.057   \\
 MTF-GLP-FS       &   &     0.097    &     3.795  &     0.092   &     0.087   &     0.132   &     0.098   &   &     0.037   &     3.515   &     0.055   &     0.083   &     0.068   &     0.056   \\
 MTF-GLP-HPM      &   &     0.102    &     3.721  &     0.106   &     0.067   &     0.109   &     0.115   &   &     0.036   &     3.557   &     0.055   &     0.077   &     0.059   &     0.058   \\
 MF               &   &     0.091    &     3.542  &     0.111   &     0.077   &     0.134   &     0.096   &   &     0.041   &     3.697   &     0.062   &     0.101   &     0.055   &     0.065   \\
 FE-HPM           &   &     0.096    &     3.786  &     0.097   &     0.077   &     0.125   &     0.099   &   &     0.039   &     3.669   &     0.059   &     0.088   &     0.064   &     0.053   \\
 SR-D             &   &     0.087    &     3.718  & \za{0.056}  &     0.268   &     0.062   &     0.305   &   & \za{0.023}  & \za{3.135}  & \za{0.035}  &     0.202   &     0.078   &     0.191   \\
 TV               &   &     0.110    &     3.576  &     0.171   &     0.693   &     0.070   &     0.114   &   &     0.056   &     4.213   &     0.100   &     0.675   &     0.051   &     0.062   \\    \cline{1-1} \cline{3-8} \cline{10-15}
 \zr PNN          &   &     0.110    &     3.858  &     0.109   &     0.471   &     0.089   &     0.115   &   &     0.046   &     3.811   &     0.069   &     0.376   &     0.032   &     0.057   \\
 A-PNN            &   &     0.103    &     3.796  &     0.096   &     0.540   &     0.095   &     0.150   &   &     0.041   &     3.791   &     0.064   &     0.497   &     0.072   &     0.075   \\
 A-PNN-FT         &   &     0.106    &     3.840  &     0.099   &     0.372   &     0.067   &     0.130   &   &     0.044   &     3.932   &     0.067   &     0.264   &     0.030   &     0.059   \\
 BDPN*            &   &     0.182    &     4.559  &     0.287   &     0.346   &     0.065   &     0.016   &   &     0.103   &     5.685   &     0.188   &     0.334   &     0.041   &     0.009   \\
 DiCNN*           &   &     0.175    &     4.306  &     0.265   &     0.342   &     0.060   &     0.030   &   &     0.147   &     5.898   &     0.263   &     0.263   & \za{0.028}  &     0.020   \\
 DRPNN*           &   &     0.144    &     3.669  &     0.240   &     0.614   &     0.148   &     0.075   &   &     0.083   &     4.702   &     0.159   &     0.502   &     0.038   &     0.041   \\
 FusionNet*       &   &     0.228    &     4.838  &     0.341   &     0.346   &     0.159   &     0.026   &   &     0.176   &     6.393   &     0.319   &     0.201   & \zb{0.036}  &     0.013   \\
 MSDCNN*          &   &     0.174    &     3.951  &     0.268   &     0.436   &     0.111   &     0.047   &   &     0.101   &     5.065   &     0.198   &     0.306   &     0.036   &     0.023   \\
 PanNet*          &   &     0.096    & \zb{3.190} &     0.129   &     0.455   &     0.076   &     0.113   &   &     0.039   &     3.785   &     0.068   &     0.356   &     0.037   &     0.071   \\
 SSQ**            &   &     0.122    &     3.792  &     0.166   &     0.335   & \zb{0.057}  &     0.052   &   &     0.055   &     4.110   &     0.099   &     0.284   &     0.036   &     0.033   \\
 GDD*             &   &     0.399    &     8.146  &     0.470   &     0.564   &     0.097   &     0.201   &   &     0.282   &    10.338   &     0.386   &     0.643   &     0.100   &     0.170   \\
 PanGan*          &   &     0.263    &     5.840  &     0.387   &     0.178   &     0.076   &     0.040   &   &     0.194   &     8.347   &     0.349   &     0.107   &     0.091   &     0.029   \\
 Z-PNN            &   & \zb{0.083}   &     3.211  &     0.155   &     0.120   &     0.133   &     0.084   &   &     0.047   &     3.868   &     0.078   &     0.092   &     0.080   &     0.038   \\    \cline{1-1} \cline{3-8} \cline{10-15}
 \zr \LPNN        &   & \za{0.043}   & \za{2.220} &     0.134   & \zb{0.054}  &     0.080   &     0.265   &   & \zb{0.026}  & \zb{3.193}  & \zb{0.049}  & \zb{0.042}  &     0.095   &     0.178   \\    \hline
\end{tabular}
\caption{Average results on GeoEye-1. Left: Genoa. Right: London+Trenton (PairMax)}
\label{tab:GE1 results}
\end{table*}

Tables~\ref{tab:WV3 results}, \ref{tab:WV2 results}, and \ref{tab:GE1 results} report numerical results for the WV3, WV2, and GE1 datasets, respectively.
For each sensor there are two test datasets, {\it e.g.}, Adelaide and Munich for WV3, for a total of 6, and
for each image, there are 6 columns of results, corresponding to three spectral and three spatial distortion metrics.
If we consider only the metrics that contribute to the proposed JESSE loss, $\DLa$, R-ERGAS, and $\DR$,
the proposed \LPNN\ (last row) ranks always first (bold) or second (underlined) among all competing methods.
This is a great achievement, and our goal from the beginning, because we think that these metrics are the most reliable indicators of quality.
Moreover, the fact that \LPNN\ succeeds in optimizing {\em both} spectral and spatial quality metrics
further confirms that these latter provide coherent indications, with no domain mismatch.
Critics will argue that this result is due, at least in part, to the alignment between the training and measurement phases.
This is probably true, so we defer any further related consideration to visual inspection.

\begin{figure}
\centering
\footnotesize
\setlength{\tabcolsep}{1mm}
\begin{tabular}{cc}
                                      MS          &                                         PAN      \\
\includegraphics[width=38mm]{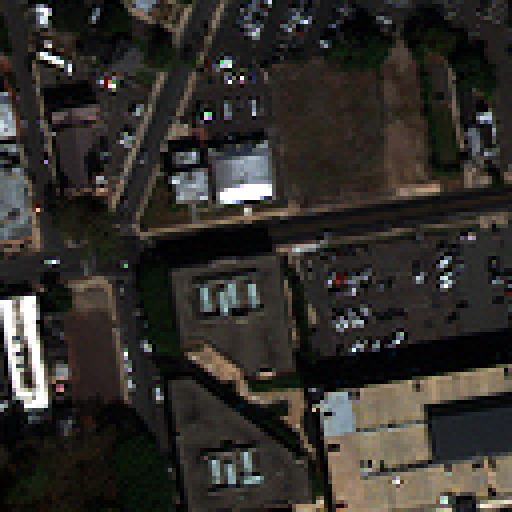}   & \includegraphics[width=38mm]{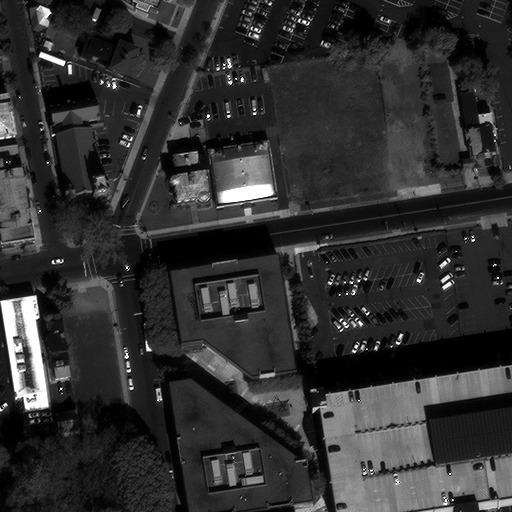} \\[2mm]
                                     \LPNN\       &                                         EXP      \\
\includegraphics[width=38mm]{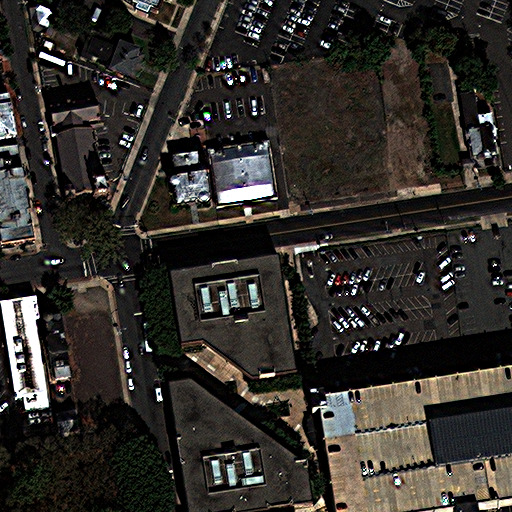} & \includegraphics[width=38mm]{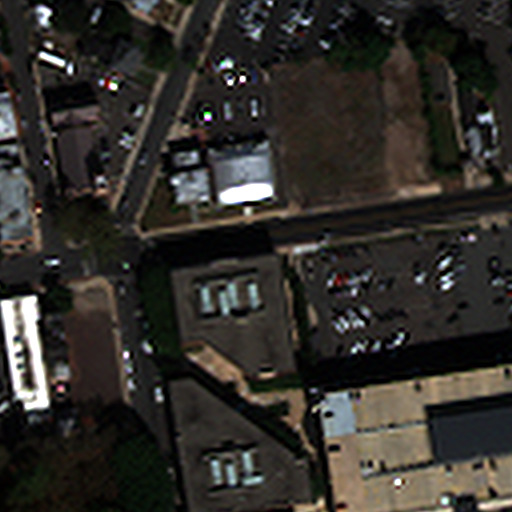} \\
\end{tabular}
\caption{
Example results for GE1 Trenton-PairMax test image.
For the \LPNN\ pansharpened image (bottom-left) we have
$\DR$=0.042, $\DS$=0.095, $\DSR$=0.178, while we have instead
$\DR$=0.805, $\DS$=0.078, $\DSR$=0.133 for the image obtained through simple interpolation of the MS (bottom-right).
Both $\DS$ and $\DSR$ provide misleading indications on spatial quality.
}
\label{fig:EXP_example}
\end{figure}

Nonetheless, even when we consider $\DL$ (third column),
\LPNN\ keeps providing very good results, typically better than all competitors except for some MRA and TV methods.
That is, spectral fidelity remains high despite the penalty given by the mismatch between MS and pansharpened output
in the presence of imperfect co-registration.
The situation is completely different when we consider $\DS$ and $\DSR$:
according to these spatial distortion indicators, \LPNN\ appears to be among the worst performers.
We are not discouraged by these results, though,
because we are skeptical on the ability of these indicators to really capture spatial quality, and report them for the sake of completeness.
Indeed, unlike for {\em spectral} quality, the problem of no-reference {\em spatial} quality assessment is far from being solved.
We refer the reader to our recent works \cite{Ciotola2022,Scarpa2022a} for a deeper analysis of these metrics and,
again, leave the final word to the visual inspection of images.
However, it is worth noting that, on some images, the proposed \LPNN\ is outperformed even by simple interpolation (EXP) according to these indicators.
This speaks volumes about their reliability.
In Fig.\ref{fig:EXP_example}, we show a relevant example that illustrates this phenomenon and fully supports our standing.

\subsection{Visual Results}

\newcommand{\ssiz}[1]{{\scriptsize #1}}
\newcommand{\best}[1]{{\scriptsize Best #1}}

\newcommand{\imAx}[1]{\includegraphics[width=19mm]{./png_comp/WV3/Adelaide/#1.jpg}}
\newcommand{\imPx}[1]{\includegraphics[width=19mm]{./png_comp/WV3/PairMax/#1.jpg}}
\newcommand{\imWy}[1]{\includegraphics[width=19mm]{./png_comp/WV2/Washington/#1.jpg}}
\newcommand{\imPy}[1]{\includegraphics[width=19mm]{./png_comp/WV2/PairMax/#1.jpg}}
\newcommand{\imGz}[1]{\includegraphics[width=19mm]{./png_comp/GE1/Genova/#1.jpg}}
\newcommand{\imPz}[1]{\includegraphics[width=19mm]{./png_comp/GE1/PairMax/#1.jpg}}

\begin{figure*}
\centering
\tiny
\setlength{\tabcolsep}{0.5mm}
\begin{tabular}{ccccccccc}
\ssiz{MS}     & \ssiz{PAN}   & \ssiz{\LPNN}       & \best{$\DLa$}      & \best{R-ERGAS}     & \best{$\DL$}         & \best{$\DR$}       & \best{$\DS$}      & \best{$\DSR$} \\[1mm]
\imAx{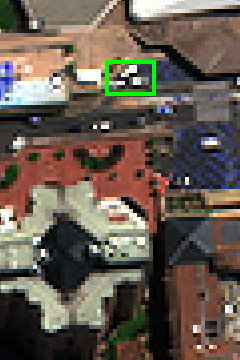}   & \imAx{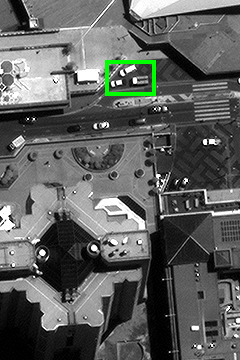} & \imAx{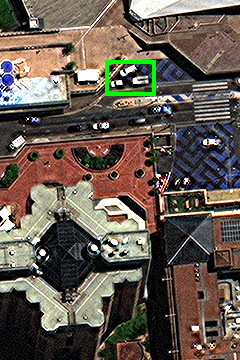} & \imAx{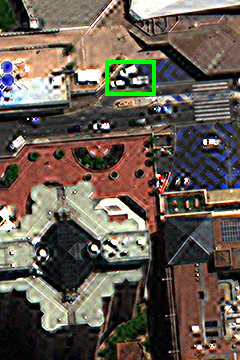}        & \imAx{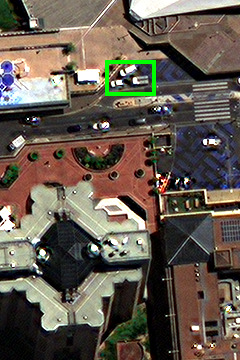} & \imAx{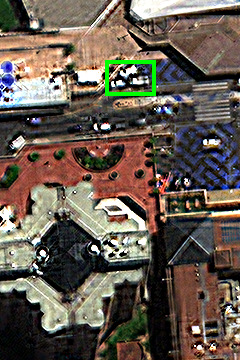}        & \imAx{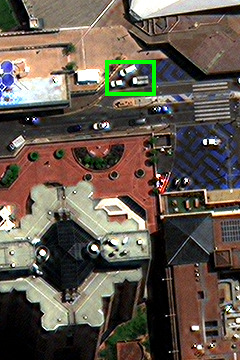}       & \imAx{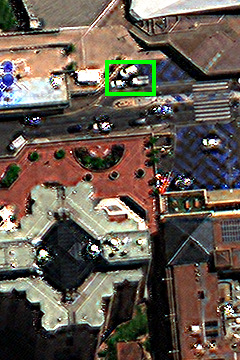}   & \imAx{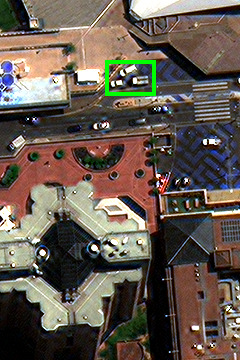} \\
              &              &      \LPNN         &       TV           &       Z-PNN        &       SR-D           &       GSA          &       PanNet      &       BT-H    \\[1mm]
\imPx{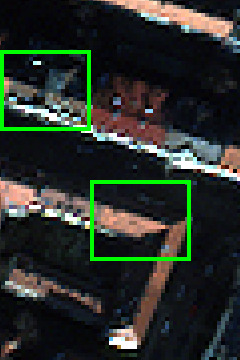}   & \imPx{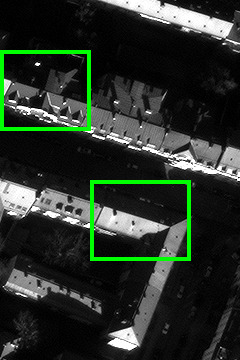} & \imPx{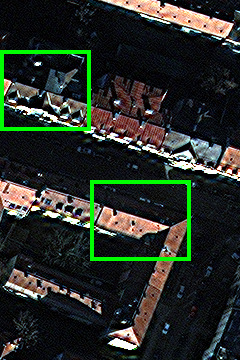} & \imPx{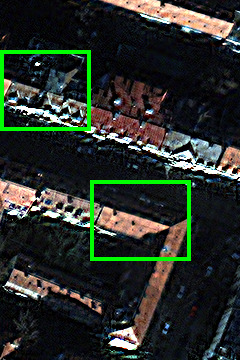}      & \imPx{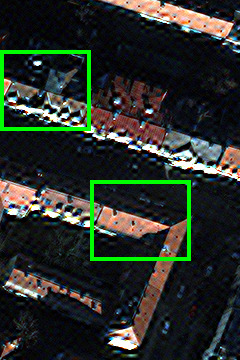}        & \imPx{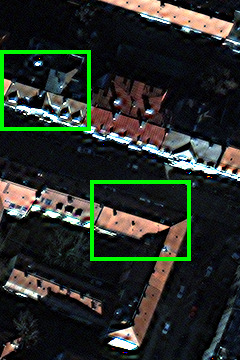}        & \imPx{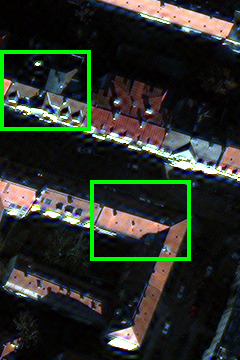}      & \imPx{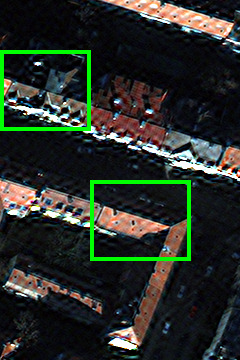} & \imPx{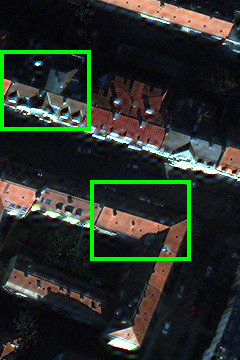}   \\
              &              &      \LPNN         &       SR-D         &       TV           &       AWLP           &       BT-H         &       A-PNN-FT    &       GS      \\
\end{tabular}
\caption{Samples from WV3 Adelaide (top) and Munich-PairMax (bottom).  Left to right: MS, PAN, \LPNN, best references.}
\label{fig:WV3_PM_crops}
\end{figure*}

\begin{figure*}
\centering
\tiny
\setlength{\tabcolsep}{0.5mm}
\begin{tabular}{ccccccccc}
\ssiz{MS}     & \ssiz{PAN}   & \ssiz{\LPNN}       & \best{$\DLa$}      & \best{R-ERGAS}     & \best{$\DL$}         & \best{$\DR$}       & \best{$\DS$}      & \best{$\DSR$}  \\[1mm]
\imWy{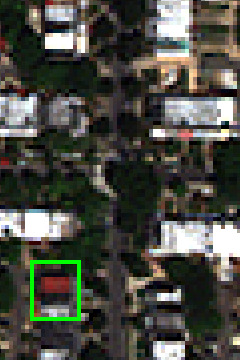}   & \imWy{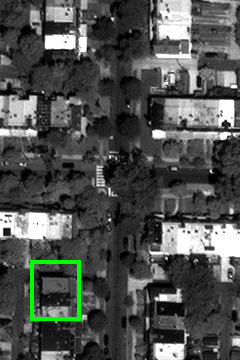} & \imWy{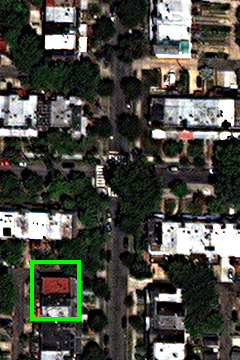} & \imWy{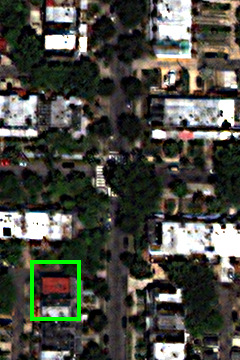}      & \imWy{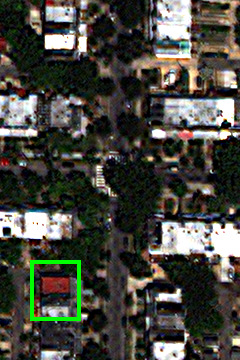}        & \imWy{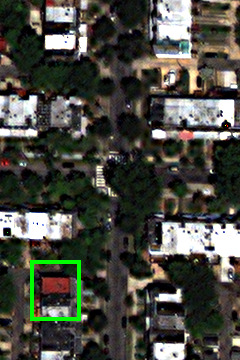}        & \imWy{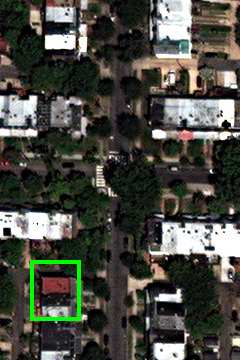} & \imWy{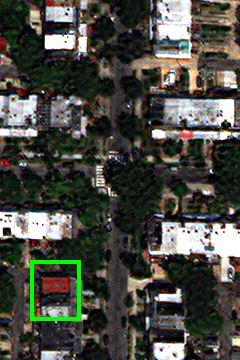} & \imWy{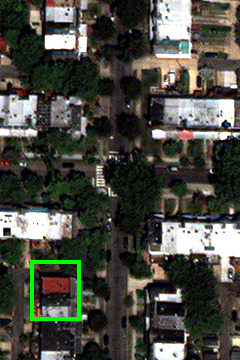}  \\
              &              &      \LPNN         &       SR-D         &       TV           &       AWLP           &       Z-PNN        &       A-PNN-FT    &       BT-H     \\[1mm]
\imPy{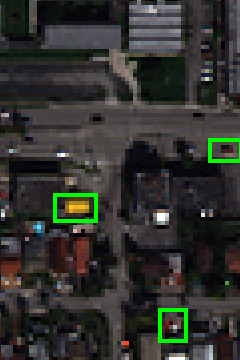}   & \imPy{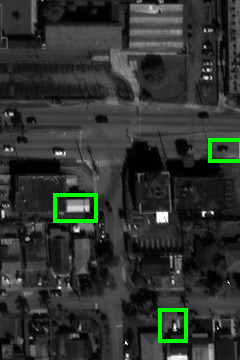} & \imPy{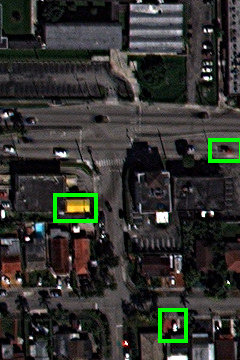} & \imPy{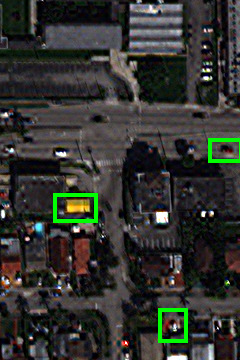}        & \imPy{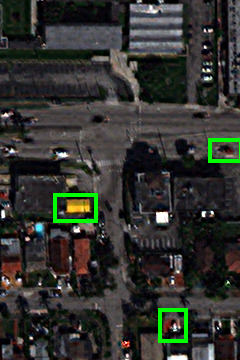}    & \imPy{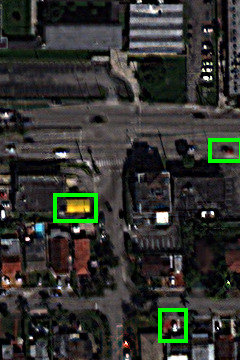}        & \imPy{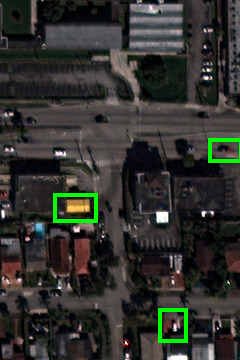} & \imPy{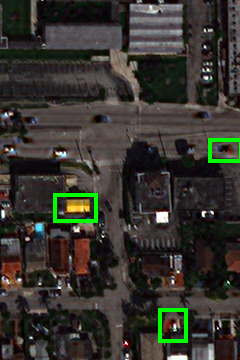}    & \imPy{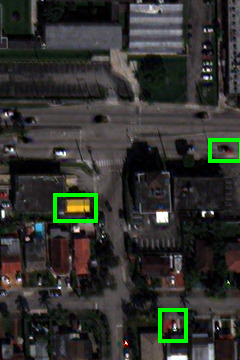}  \\
              &              &      \LPNN         &       TV           &       PanNet       &       SR-D           &       Z-PNN        &       DRPNN       &       BT-H     \\
\end{tabular}
\caption{Samples from WV2 Washington (top) and Miami-PairMax (bottom). Left to right: MS, PAN, \LPNN, best references.}
\label{fig:WV2_PM_crops}
\end{figure*}

\begin{figure*}
\centering
\tiny
\setlength{\tabcolsep}{0.5mm}
\begin{tabular}{ccccccccc}
\ssiz{MS}     & \ssiz{PAN}   & \ssiz{\LPNN}       & \best{$\DLa$}      & \best{R-ERGAS}     & \best{$\DL$}         & \best{$\DR$}       & \best{$\DS$}      & \best{$\DSR$}  \\[1mm]
\imGz{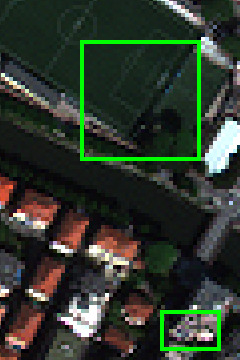}   & \imGz{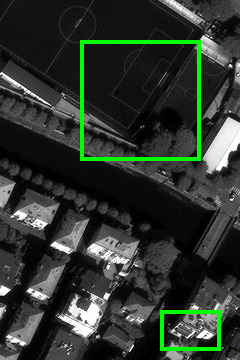} & \imGz{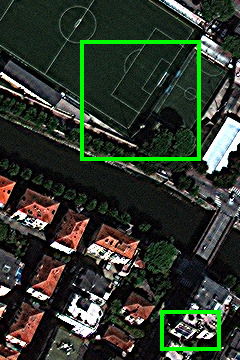} & \imGz{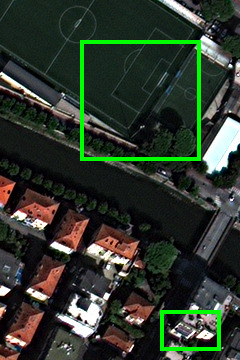} & \imGz{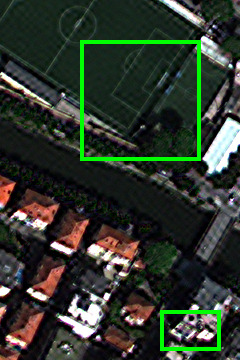}    & \imGz{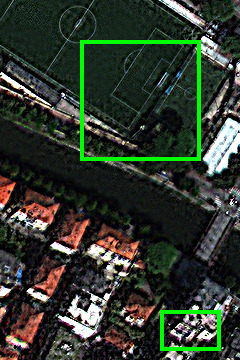}        & \imGz{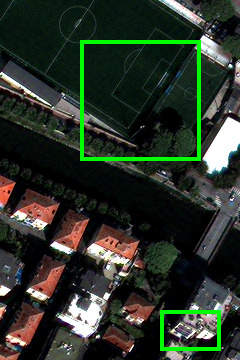}       & \imGz{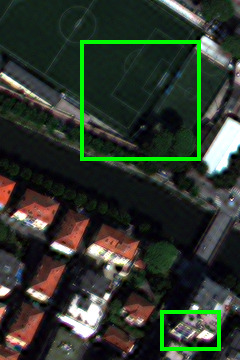}    & \imGz{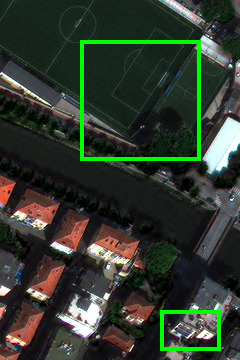}    \\
              &              &      \LPNN         &       Z-PNN        &       PanNet       &       SR-D           &       GSA          &       PRACS       &       GS       \\[1mm]
\imPz{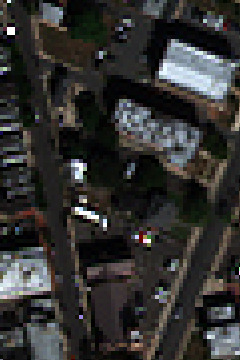}   & \imPz{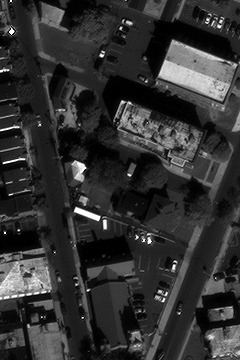} & \imPz{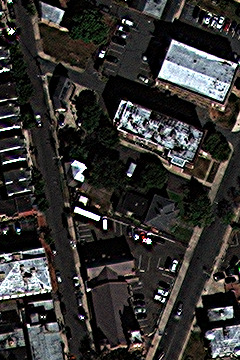} & \imPz{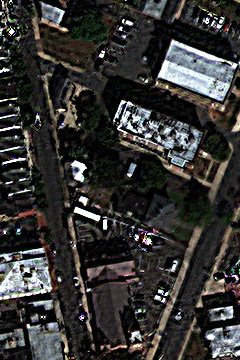}      & \imPz{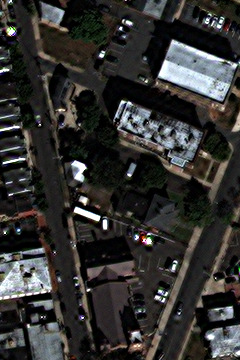}   & \imPz{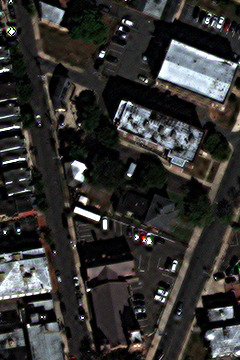} & \imPz{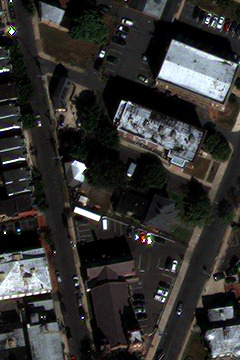}       & \imPz{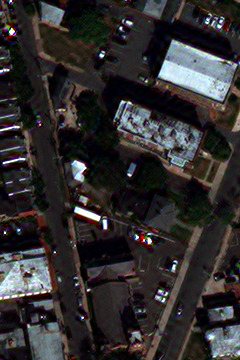}   & \imPz{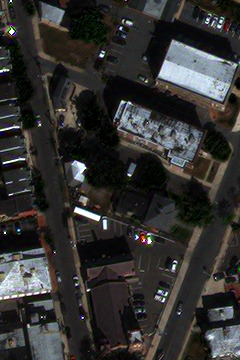}    \\
              &              &      \LPNN         &       SR-D         &       MTF-GLP      &       MTF-GLP-HPM    &       GSA          &       DiCNN       &       GS       \\
\end{tabular}
\caption{Samples from GE1 Genoa (top) and Trenton-PairMax (bottom).    Left to right: MS, PAN, \LPNN, best references.}
\label{fig:GE1_PM_crops}
\end{figure*}

As already mentioned,
finding a reliable measure of pansharpened image quality is still under intense research and there is no consensus, to date,
on which metric or combination of metrics best fits the judgment and needs of end users.
Therefore, we rely on visual inspection to confirm (or possibly disconfirm) our conclusions.
An expert viewer can still spot spectral and spatial artifacts better than compact (averaged) indicators can.

Figures \ref{fig:WV3_PM_crops}, \ref{fig:WV2_PM_crops}, and \ref{fig:GE1_PM_crops} show visual results for the WV3, WV2, and GE1 datasets, respectively.
For each sensor, two crops are selected, one for each dataset, {\it e.g.}, Adelaide and Munich-PairMax for WV3.
For each selected crop,
we show original MS and PAN, on the left,
followed by the pansharpened outputs of the proposed method, \LPNN, and the six strongest competitors, one per each distortion metric.
For example, next to \LPNN, we show the image generated by the method that performs best (except for \LPNN\ itself) under the $\DLa$ distortion metric.
This choice allows us to limit the number of images to inspect,
but also to assess the proposed method against the most competitive reference methods.
Note that the ``challenger'' under a given metric changes from image to image.
For example, for the WV3 test set and $\DLa$, it is TV for Adelaide but SR-D for Munich.
Of course, if a method is the best competitor under more than one metric, we avoid showing the same image multiple times and use the next in list.
Multispectral images are shown using an RGB (red-green-blue bands) composition.
In theory, artifacts may happen to occur {\em only} in bands not shown, thus evading visual inspection.
However, we checked several images, characterized by both low and high spectral distortion, and never spotted such phenomena.

\begin{figure*}[htbp]
\centering
\includegraphics[scale=0.85]{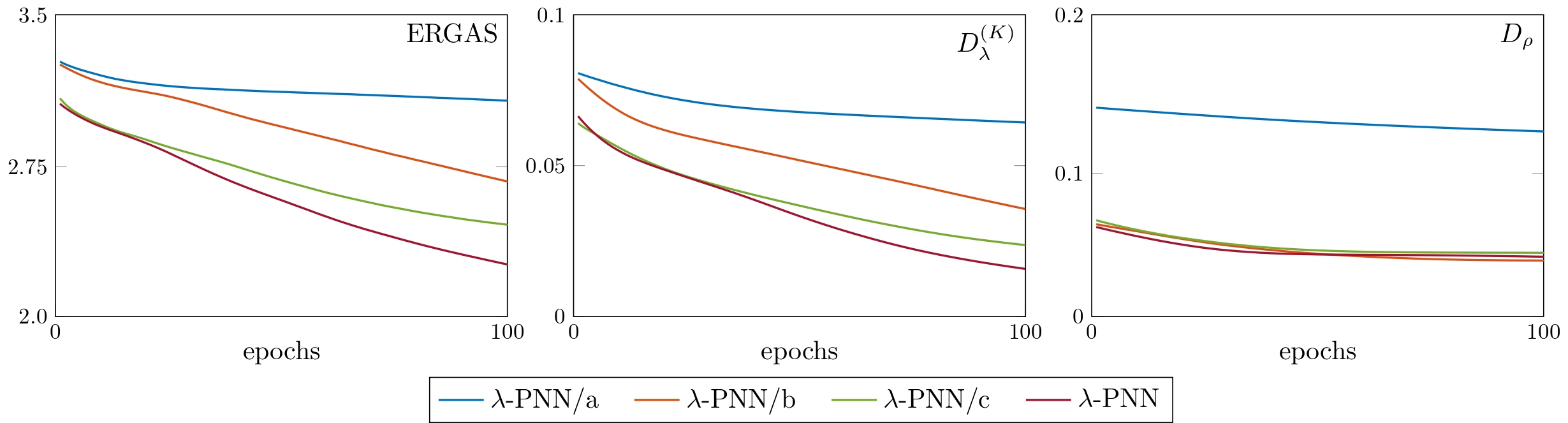}
\caption{Loss curves observed during target adaptation to a GE1 Genoa image. All models pre-trained on the GE1 training set.}
\label{fig:network_effects}
\end{figure*}

To begin, we study the absolute quality of the images pansharpened by \LPNN,
that is, whether they can be satisfactory for the end user, no matter how they compare with competitors.
To this end, we analyze PAN, MS, and the {\LPNN} image together, looking for possible spectral or spatial artifacts.
With fastidious scrutiny, we singled out a few geometric errors (marked by green boxes for easier inspection), like in the upper-most truck in Adelaide,
or in the facade of the bottom-right building in Genoa.
Likewise, minor color deviations can sometimes be observed with respect to the MS, {\it e.g.,} some very small cars in Miami.
All in all, these are minor exceptions to the general rule of high spatial and spectral fidelity,
with textures accurately copied from PAN and colors closely matching those of MS.
Nonetheless, we notice an annoying quasi-periodic pattern in some surfaces supposed to be flat, like the streets of Miami.
However, a close inspection reveals that such patterns were already present in the PAN image.
\LPNN, by forcing a strong correlation with the PAN, replicates such patterns which are made more visible by the injection of color.
In general, the quality of \LPNN\ images appears to be fully satisfactory and to meet the original goal of high joint spatial-spectral fidelity.

Let us now proceed to a comparative analysis,
focusing on the products that appear to be the most competitive under one or more distortion metrics.
First, we consider the methods that provide the least spectral distortion under the $\DLa$, R-ERGAS, and $\DL$ metrics.
For WV3 Adelaide these are TV, MF, and SR-D.
They all show a significant loss of resolution with respect to \LPNN\ and to the PAN,
somewhat milder for SR-D which, however, exhibits spatial artifacts.
On the other hand, for all of them, $\DR$ is much larger than for \LPNN\, signaling a limited correlation with the PAN.
What is observed for Adelaide, repeats over and over as we go through the other test images, and always in agreement with numerical results.
In all cases, good spectral quality is obtained at the price of reduced resolution or spatial artifacts,
Therefore, it is fair to conclude that only the proposed \LPNN\ seems able to
consistently provide {\em both} good spectral and good spatial quality.

Similar conclusions can be drawn by looking at the best competitors under the spatial distortion metrics.
In this case, we discard right away the last two columns because,
as already noted, the $\DS$ and $\DSR$ indicators are not very reliable and often point to images with unsatisfactory spatial quality.
Consider for example the PRACS image obtained for Genoa (Fig.10), which has the best $\DS$ but shows a dramatic loss of resolution.
We focus instead on the best $\DR$ images, which are always characterized by good spatial quality.
Spectral quality is generally good for them, but deeper scrutiny reveals several problematic details.
A few examples (marked again by green boxes):
BT-H (Munich) is generally bluish, especially in high-contrast areas;
Z-PNN (Washington and Miami) has rooftops with desaturated colors, and a yellow rooftop (Miami) with plain wrong hue;
the football field of GSA (Genoa) is too dark.
Again, \LPNN\ appears to provide the fewest spatial and spectral artifacts.

\subsection{Ablation studies: architecture}

{\renewcommand{\ru}{\rule{0mm}{3mm}}
\begin{table}
\centering
\setlength{\tabcolsep}{2pt}
\begin{tabular}{c|cccc}
\hline
\ru architecture  & ConvLayers & ResBlocks & CBAM & R-CBAM \\ \hline
\ru \LPNN/a       &   \checkmark & --         & --         &  --        \\
\ru \LPNN/b       &   \checkmark & \checkmark & --         &  --        \\
\ru \LPNN/c       &   \checkmark & \checkmark & \checkmark &  --        \\
\ru full \LPNN    &   \checkmark & \checkmark & --         & \checkmark \\
\hline
\end{tabular}
\caption{\LPNN\ versions considered in the ablation study, from baseline, with just 3 convolutional layers, to full fledged.}
\label{tab:architectures}
\end{table}
}

To gain further insight into the effectiveness of our design choices, we carry out some ablation studies, starting with architecture.
To this end, we compare four versions of the proposed model (see Tab.\ref{tab:architectures}),
starting with a basic CNN and gradually adding new layers until we get the full \LPNN\ architecture of Fig.\ref{fig:network}(a). The basic model comprises just 3 convolutional layers,
then, in \LPNN/b the 2 ResBlocks (red boxes in are added,
followed by the 2 attention layers, conventional (CBAM) in \LPNN/c or residual (R-CBAM) in full \LPNN.

\newcommand{\imLo}[1]{\includegraphics[width=1.7cm]{./png_losses/#1.jpg}}
\begin{figure*}
\centering
\scriptsize
\setlength{\tabcolsep}{0.5mm}
\begin{tabular}{cccccccccc}
      MS      &       PAN  &       JESSE      & w/o $\DLa$      &    w/o ERGAS    & $\DSR \to \DR$      &       SSQ         &       GDD         &       PG         &       Z-PNN     \\ [1mm]
\imLo{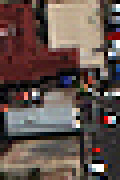}     & \imLo{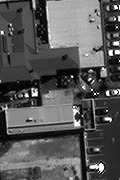} & \imLo{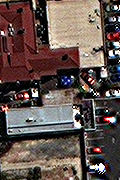} & \imLo{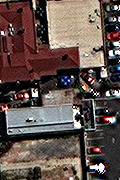} & \imLo{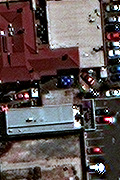} & \imLo{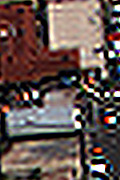}   & \imLo{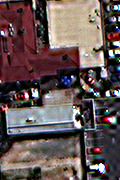} & \imLo{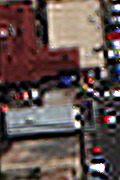} & \imLo{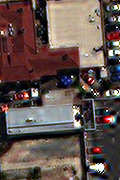} & \imLo{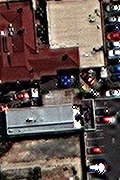} \\
\end{tabular}
\caption{Sample from WV3 X-Val (Adelaide). Left to right: MS, PAN, pansharpened by \LPNN\ with various losses.}
\label{fig:loss_WV3_crop}
\end{figure*}

In Fig.\ref{fig:network_effects} we show the evolution of the three selected metrics during the target adaptation phase.
Models were all pre-trained in the same conditions on the GE1 training set and then adapted to the target Genoa image (GE1 X-Val dataset).
The initial point of each curve corresponds to the distortion observed with pre-trained weights,
which is then gradually reduced with target adaptation.
The basic model (\LPNN/a, blue curves), is clearly unable to exploit the potential of full-resolution training.
Both spatial and spectral distortions are the largest and, what is worse, they do not decrease significantly with fine-tuning.
We conjecture that with just three layers, the composite receptive field is too small to capture necessary dependencies.
All deeper architectures perform much better.
By including residual blocks (\LPNN/b, orange) the spatial distortion $\DR$ is more than halved, even with the pre-trained weights.
However, the spectral distortion remains relatively large and reduces somewhat only after intense fine-tuning.
The further inclusion of attention mechanisms seems to unlock the potential of full-resolution training.
With CBAM blocks (\LPNN/c, green) and even more with R-CBAM (\LPNN, red) a major boost in spectral quality is obtained.
Eventually,
\LPNN\ reduces all distortion metrics with respect to the basic reference, by as much as 20\% (R-ERGAS), 40\% ($\DLa$), and 60\% ($\DR$).

It is also worth underlining that target adaptation plays a pivotal role in achieving such good results,
since it keeps reducing the spectral distortion during the whole process and, apparently, could provide further gains with more epochs.
All this is possible thanks to the proposed fast target adaptation procedure.
Otherwise, with large images and large models, only a few epochs would have been affordable.

\subsection{Ablation studies: loss function}

We now turn to assess the impact of the proposed JESSE loss on the good performance of \LPNN.
Therefore, we now keep the architecture fixed and change only the loss function.
For each of these losses, we train anew \LPNN\ on the VW3 training set.
Numerical results on the WV3 X-Val dataset (24 tiles from the Adelaide image) are reported in Tab.\ref{tab:loss_comparison},
considering only the $\DL$ and $\DR$ metrics for brevity.
In the upper part of the table, we report results for the JESSE loss itself (first row) and some ablated variants:
without the $\DLa$ spectral (sub)term, without the R-ERGAS spectral (sub)term, with $\DSR$ in place of $\DR$ as spatial term.
In the lower part, instead, we show the results obtained
when the JESSE loss is replaced altogether by one of the loss functions proposed for other pansharpening methods that rely on unsupervised full-resolution training,
SSQ, GDD, PG, and Z-PNN.
Together with the numerical results, we rely also on a visual example.
In Fig.\ref{fig:loss_WV3_crop}, for a small test crop we show, as usual, MS and PAN original data,
followed by the images pansharpened by \LPNN\ trained with each of the aforementioned losses.
Note that the automatic alignment mechanism is deactivated in all cases to ensure a fair comparison.

By keeping the JESSE loss, but removing one of the spectral subterms (second and third row),
the balance between spectral and spatial fidelity changes somewhat in favor of the latter, but not dramatically so.
However, the pansharpened samples show some spectral aberrations.
This is especially marked when the ERGAS term is removed, with clear hue distortions ({\it e.g.}, the yellow rooftop on the top-right).
Instead, replacing the correlation-based metric, $\DR$, with the popular $\DSR$ metric in the spatial loss term (row 4),
has a catastrophic impact on image quality, as clear from both the numbers and the sample image.
On the other hand, this is only a further confirmation of what was observed before.

Let us now turn to consider completely different loss functions, copied by recently published papers.
We keep using the \LPNN\ architecture and replace only spatial and spectral loss terms with those proposed in the references.
With this study, we want to understand whether the good performance of \LPNN\ is mainly due to its improved architecture
or the loss function plays a fundamental role in it.
Both the numbers reported in the bottom part of the table, (rows 5 to 8) and the pansharpened images provide unambiguous indications.
With all other losses, except for Z-PNN, a significant quality impairment is observed, especially in the spatial resolution,
as also confirmed by the $\DR$ indicator, sometimes very high.
With the Z-PNN loss, instead, the quality impairment is minimal both visually and according to the quality indicators (about 10\% worse).
On the other hand, this is the ancestor of the JESSE loss and exhibits already most of its good qualities.

\begin{table}
\centering
\setlength{\tabcolsep}{3mm}
\begin{tabular}{l|c|l|cc} \hline
\multicolumn{3}{c}{\ru Loss function}                                  & $\DL$   & $\DR$   \\ \hline
1) &                        & \ru JESSE as is                          &  0.041  &  0.040  \\
2) & variants               & \ru w/o $\DLa$                           &  0.055  &  0.034  \\
3) & of JESSE               & \ru w/o ERGAS                            &  0.071  &  0.028  \\
4) &                        & \ru $\DSR \to \DR$                       &  0.013  &  0.768  \\ \hline
5) &                        & \ru $\LL_{\rm SSQ}$ \cite{Luo2020}       &  0.065  &  0.268  \\
6) & from other             & \ru $\LL_{\rm GDD}$ \cite{Uezato2020}    &  0.067  &  0.961  \\
7) & papers                 & \ru $\LL_{\rm PG}$ \cite{Ma2020}         &  0.100  &  0.110  \\
8) &                        & \ru $\LL_{\rm Z-PNN}$ \cite{Ciotola2022} &  0.045  &  0.044  \\ \hline
\end{tabular}
\caption{Average results on WV3 X-Val (Adelaide)}
\label{tab:loss_comparison}
\end{table}

\subsection{Ablation studies: target adaptation}

\begin{figure}
\centering
\includegraphics[width=0.75\columnwidth]{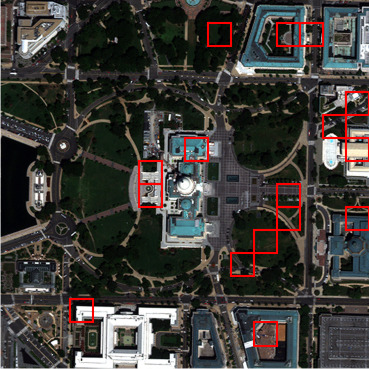}
\caption{Example of tile selection for fast target-adaptation.}
\label{fig:k-means}
\end{figure}

In our experiments, we used always the proposed fast target adaptation (TA) procedure.
For the large 2048$\times$2048-pixel test images, based on preliminary experiments, we used 16 tiles of 256$\times$256-pixels.
In Fig.\ref{fig:k-means} we show one such image together with 16 tiles selected for fine-tuning, representative of the different contexts and spatial layouts observed in the scene.
For the same image, in Tab.\ref{tab:fast_ta_metrics} we report numerical results observed with \LPNN\ pansharpening.
In the upper part of the table, we compare
fast TA with conventional TA carried out on the whole image, and with the case where no TA is used.
For each solution, we report the processing time and the spectral and spatial distortions, measured by $\DLa$ and $\DR$.
It is clear that the fast procedure has little or no impact on quality indicators while sharply reducing the processing time,
from almost 15 minutes to about 40 seconds, certainly viable for real-world applications.
Of course, this extra time is still much larger that the 2.2 second inference time.
However, the choice of not performing TA at all is not appealing, as this would significantly worsen both spectral and spatial indicators.

In the lower part of the table we study how performance depends on the choice of parameters,
showing a number of combinations of tile size (from 512$\times$512 to 128$\times$128 pixels) and number of tiles (from 64 to 4)
going from more conservative to more aggressive.
It appears that, barring the extreme case of just 4 small tiles,
distortion indicators are not significantly affected by Fast TA, with some impairments observed mainly on spectral quality.
Eventually, we selected the parameters that ensure the least increase in spectral distortion, despite the larger processing time.
In summary,
with a judicious choice of the tuning dataset, target adaptation becomes not only effective, ensuring good generalization,
but also efficient enough to allow its application with any image and model.

{\renewcommand{\ru}{\rule{0mm}{2.8mm}}
\begin{table}
\centering
\setlength{\tabcolsep}{6pt}
\begin{tabular}{cc|r|cc}
\multicolumn{2}{c|}{Target Adaptation}         &   \ru Time (s) &   $\DLa$   &    $\DR$   \\ \hline
\multicolumn{2}{c|}{fast ($256\times256, 16$)} &   \ru  37.6~   &   0.0141   &   0.0381   \\ \hline
\multicolumn{2}{c|}{conventional}              &   \ru 824.6~   &   0.0137   &   0.0396   \\ \hline
\multicolumn{2}{c|}{no t.a.}                   &   \ru   0.0~   &   0.0193   &   0.0503   \\ \hline \\
tile size                     & $\#$ tiles     &                &            &            \\ \hline
 $512\times512$               &  4~~           &   \ru  68.6~   &   0.0140   &   0.0386   \\ \hline
                              & 16~~~          &   \ru  37.6~   &   0.0141   &   0.0381   \\
 $256\times256$               &  8~~           &   \ru  24.2~   &   0.0144   &   0.0376   \\
                              &  4~~           &   \ru  18.0~   &   0.0143   &   0.0382   \\ \hline
                              & 64~~~          &   \ru  31.9~   &   0.0149   &   0.0354   \\
                              & 32~~~          &   \ru  18.5~   &   0.0145   &   0.0371   \\
 $128\times128$               & 16~~~          &   \ru  12.1~   &   0.0150   &   0.0361   \\
                              &  8~~           &   \ru   7.7~   &   0.0148   &   0.0372   \\
                              &  4~~           &   \ru   5.7~   &   0.0155   &   0.0379   \\ \hline
\hline
\end{tabular}
\caption{Results of sample experiments with conventional and fast target adaptation on a 2048$\times$2048 test image.}
\label{tab:fast_ta_metrics}
\end{table}
}

\section{Conclusions}
We proposed a new deep learning-based pansharpening model, \LPNN, with unsupervised training on full-resolution data.
Besides architectural improvements, with spatial and spectral attention modules,
a key asset of the proposal is the JESSE loss, which jointly promotes the spectral and spatial fidelity of the output images to the available reference data.
Moreover, we proposed a fast target adaptation procedure to ensure good generalization ability in all practical applications.
To assess the proposed method we performed a large number of experiments on images acquired by various sensors,
obtaining always excellent numerical results and convincing quality of output images.

We believe that full-resolution unsupervised training offers the best opportunity to unlock the full potential of deep learning in pansharpening.
However, despite the good results obtained with the proposed method, there are still problems to be solved and room for improvement.
An immediate goal is to characterize and ultimately eliminate the anomalous patterns observed in some cases.
More ambitious and long-term objectives concern the extension of the method to some challenging cases:
{\it  i)} the joint co-registration and pansharpening of images with moving objects, {\it e.g.}, vehicles, which cause long-range local shifts;
{\it ii)} the pansharpening of multi- or hyper-spectral bands weakly correlated with the PAN, for which a new suitable spatial loss term needs to be defined.

\balance
\bibliographystyle{IEEEtran}
\bibliography{refs}

\end{document}